\documentclass[%
 reprint,
 superscriptaddress,
 amsmath,amssymb,
 aps,
prb,
floatfix,
]{revtex4-1}

\usepackage{graphicx}
\usepackage{dcolumn}
\usepackage{bm}
\usepackage[normalem]{ulem}
\usepackage{hyperref}
\hypersetup{colorlinks = true, citecolor = blue, breaklinks = true}

\usepackage[usenames,dvipsnames]{xcolor}
\usepackage{textcomp}
\usepackage{multirow}
\usepackage{xfrac}
\graphicspath{{../images_v3/}}

\newcommand{\angstrom}{\mbox{\normalfont\AA}}

\newcommand{\editor}[2]{%
  \expandafter\newcommand\csname #1note\endcsname[1]{%
    \textcolor{#2}{(\textbf{#1:} ##1)}}%
  \expandafter\newcommand\csname #1\endcsname[1]{%
    \textcolor{#2}{##1}}%
  \expandafter\newcommand\csname #1cancel\endcsname[1]{%
    \textcolor{#2}{\sout{##1}}}%
  \expandafter\newcommand\csname #1change\endcsname[2]{%
    \textcolor{#2}{\sout{##1} ##2}}%
  \newenvironment{#1text}{\color{#2}}{\color{black}}
}
\editor{CRSI}{orange}
\editor{CRE}{violet}
\editor{IT}{blue}

\usepackage{mhchem}

\usepackage[resetlabels]{multibib}
\newcites{SI}{References}

\begin{document}

\title{Self-consistent DFT+$U$+$V$ study of oxygen vacancies in SrTiO$_3$}

\affiliation{%
Department of Chemistry and Biochemistry and National Centre for Computational Design and Discovery of Novel Materials (MARVEL), University of Bern, Freiestrasse 3, CH-3012 Bern, Switzerland 
}%
\affiliation{%
Theory and Simulation of Materials (THEOS) and National Centre for Computational Design and Discovery of Novel Materials (MARVEL), Ecole Polytechnique F\'ed\'erale de Lausanne, CH-1015 Lausanne, Switzerland
}%
\affiliation{%
Department of Physics, University of Pavia, Via A. Bassi 6, 27100 Pavia, Italy.
}%

\author{Chiara Ricca}
\affiliation{%
Department of Chemistry and Biochemistry and National Centre for Computational Design and Discovery of Novel Materials (MARVEL), University of Bern, Freiestrasse 3, CH-3012 Bern, Switzerland 
}%

\author{Iurii Timrov}
\affiliation{%
Theory and Simulation of Materials (THEOS) and National Centre for Computational Design and Discovery of Novel Materials (MARVEL), Ecole Polytechnique F\'ed\'erale de Lausanne, CH-1015 Lausanne, Switzerland
}%

\author{Matteo Cococcioni}
\affiliation{%
Theory and Simulation of Materials (THEOS) and National Centre for Computational Design and Discovery of Novel Materials (MARVEL), Ecole Polytechnique F\'ed\'erale de Lausanne, CH-1015 Lausanne, Switzerland
}%
\affiliation{%
Department of Physics, University of Pavia, Via A. Bassi 6, 27100 Pavia, Italy.
}%

\author{Nicola Marzari}
\affiliation{%
Theory and Simulation of Materials (THEOS) and National Centre for Computational Design and Discovery of Novel Materials (MARVEL), Ecole Polytechnique F\'ed\'erale de Lausanne, CH-1015 Lausanne, Switzerland
}%

\author{Ulrich Aschauer}
\email{ulrich.aschauer@dcb.unibe.ch}
\affiliation{%
Department of Chemistry and Biochemistry and National Centre for Computational Design and Discovery of Novel Materials (MARVEL), University of Bern, Freiestrasse 3, CH-3012 Bern, Switzerland 
}%

\date{\today}

\begin{abstract}
Contradictory theoretical results for oxygen vacancies in SrTiO$_3$ (STO) were often related to the peculiar properties of STO, which is a $d^0$ transition metal oxide with mixed ionic-covalent bonding. Here, we apply, for the first time, density functional theory (DFT) within the extended Hubbard DFT+$U$+$V$ approach, including on-site as well as inter-site electronic interactions, to study oxygen-deficient STO with Hubbard $U$ and $V$ parameters computed self-consistently via density-functional perturbation theory. Our results demonstrate that the extended Hubbard functional is a promising approach to study defects in materials with electronic properties similar to STO. Indeed, DFT+$U$+$V$ provides a better description of stoichiometric STO compared to standard DFT or DFT+$U$, the band gap and crystal field splitting being in good agreement with experiments. In turn, also the description of the electronic properties of oxygen vacancies in STO is improved, with formation energies in excellent agreement with experiments as well as results obtained with the most frequently used hybrid functionals, however at a fraction of the computational cost. While our results do not fully resolve the contradictory findings reported in literature, our systematic approach leads to a deeper understanding of their origin, which stems from different cell sizes, STO phases, the exchange-correlation functional, and the treatment of structural relaxations and spin-polarization.
\end{abstract}

\maketitle

\section{\label{sec:intro}Introduction}

\begin{figure}
 \centering
 \includegraphics[width=0.95\columnwidth]{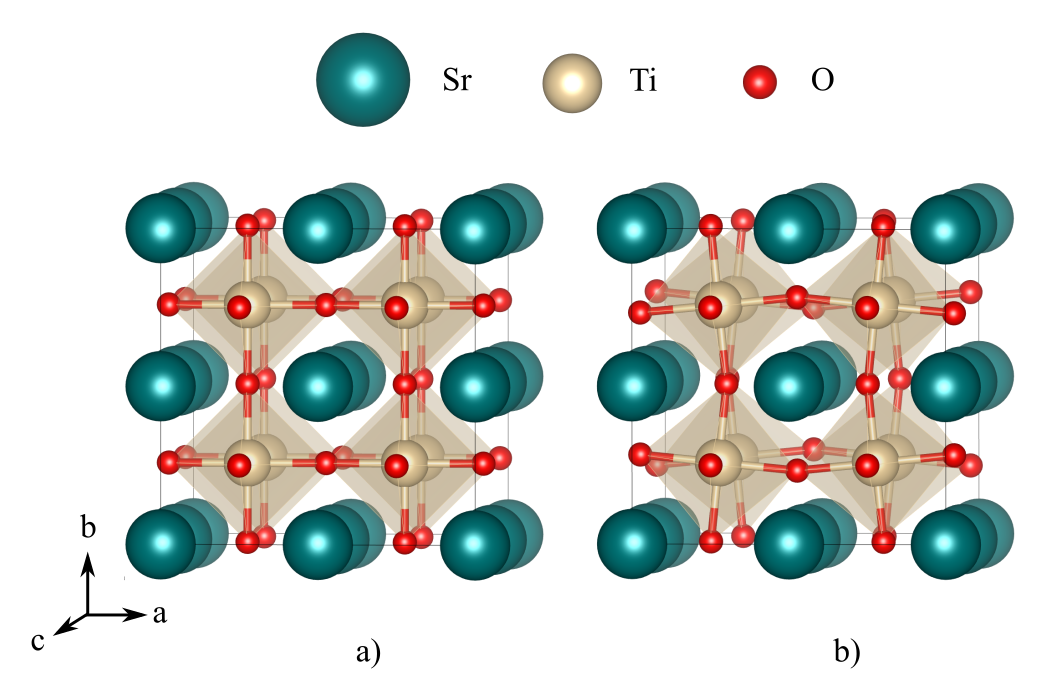}
 \caption{(2$\times$2$\times$2) supercell of stoichiometric a) cubic and b) antiferrodistortive \ce{SrTiO3}.}
\label{fig:STO_structure_bulk}
\end{figure}

\ce{SrTiO3} (STO) is a perovskite oxide that around 105~K~\cite{Shirane1969, COWLEY1969181} undergoes a transition from a high-temperature cubic (space group $Pm\bar{3}m$, Fig.~\ref{fig:STO_structure_bulk}a) to a tetragonal antiferrodistortive phase (AFD, space group $I4/mcm$, Fig.~\ref{fig:STO_structure_bulk}b) in which \ce{TiO6} octahedra rotate around the $c$-axis with out-of-phase rotations in consecutive layers ($a^0a^0c^-$ in Glazer notation\cite{Glazer:1972eb}). Along with \ce{BaTiO3}, \ce{CaTiO3}, and \ce{PbTiO3}, STO is often considered as a prototypical perovskite material and a lot of research has been dedicated to understanding its properties. Unlike \ce{BaTiO3}, STO is not ferroelectric as the condensation of the computationally predicted ferroelectric soft-mode~\cite{Zhong1996, Vanderbilt16998STO} is prevented by quantum fluctuations even at the lowest reachable temperatures~\cite{Muller1979}.

Defects and doping can be used to tune the functional properties of perovskite oxides and to induce new physics in these materials. In particular, oxygen vacancies (V$_\mathrm{O}$) were found to result in rich variations of the physical properties of STO, causing an insulator-to-metal transition and n-type conductivity\cite{Calvani1993, Moos1997, Ohtomo2007, Lee1971}, changing the optical emission properties~\cite{Kan2005, Wang2009} and having an important effect on the transition between the AFD and the cubic phase~\cite{Hunnefeld2002,Buban2004}. Understanding the properties of oxygen vacancies is therefore a crucial prerequisite to establish the physics of this material. Despite many experimental~\cite{Calvani1993, Moos1997, Schooley1964, Koonce1967, PHILLIPS1969356, Binnig1980, Wang2009, Hunnefeld2002, Ohtomo2007, Kan2005, Shibuya2007, Zhang2008, Kim2009, Tufte1967, GONG1991320, Lee1971, Longo2008, Schwarz01051975} and theoretical~\cite{Shanthi1998, djermouni20109, Jeschke2015, Cuong2007, Astala200181, Tanaka2003, Luo2004, Cuong2007, Mitra2012, Choi2013, Kim2009, Hou2010, Lin2012, Evarestov2006, Alexandrov2009, Carrasco2005, Zhukovskii2009, ZHUKOVSKII20091359, Buban2004, Astala2001, ZHANG20121770, Ricci2003} investigations, the nature of V$_\mathrm{O}$ defects in STO still remains unclear due to contradictory results, especially related to the electronic states associated with oxygen vacancies as will be reviewed in Sec. \ref{sec:review}.
The ambiguity associated with defect states in STO was often explained with the peculiar electronic properties of the material. The mixture of very ionic Sr--O and mixed ionic-covalent Ti--O bonds together with crystal-field effects of the Ti 3$d$ orbitals lead to a competition between trapping the two extra electrons left in the lattice upon defect formation in the vacancy (F center) and localizing them on Ti-$3d$ orbitals~\cite{Ricci2003, Carrasco2005, Evarestov2006, Usvyat2004, Hamid2009}.

Due to self-interaction errors (SIE), standard density-functional theory (DFT), based on the local-density (LDA) or generalized-gradient (GGA) approximation, often fails to predict the electronic properties of materials containing transition-metal or rare-earth elements with localized $d$ or $f$ states. Hybrid DFT, incorporating a fraction of exact exchange into standard DFT functionals, could be used to correct for SIE. However, the application of hybrid functionals is still quite expensive for point-defect DFT calculations that generally require large supercells. Furthermore, the fraction of exact exchange is a material dependent parameter~\cite{Franchini2014}. In practice, this parameter is determined empirically by fitting to experimental properties, or, more often, is kept fixed at the default value suggested for each hybrid functional, thus disregarding the material dependence. In the latter case, overestimated band gaps are usually obtained for STO~\cite{Carrasco2005, Buban2004, Evarestov2006, Mitra2012}, whereas in the former case, it was shown that the fraction of exact exchange strongly depends on the experimental property used for fitting~\cite{Franchini2014, Zhang2005hybrid, onishi2008hybrid}. Due to their modest computational cost and  intuitive physical picture, DFT+$U$ approaches~\cite{anisimov1991band, anisimov1997first, Dudarev1998} are a popular alternative to treat transition-metal systems with improved accuracy compared to standard LDA and GGA functionals. This Hubbard functional augments the standard DFT energy by a Coulomb repulsion term between strongly localized $d$ or $f$ electrons belonging to the same Hubbard atom. This correction is proportional to the occupation numbers of atomic states on the given site, multiplied with the strength of the interaction, which is determined by the ``on-site'' Hubbard $U$ parameter~\cite{anisimov1991band, anisimov1997first, dudarev1998electron}.

STO, however, has a nominal $d^0$ configuration and is a band insulator rather than a correlated Mott insulator. Localization of electrons on $d$ states should therefore play a minor role, while inter-site hybridization is expected to be more important. DFT+$U$+$V$~\cite{Campo2010} is an extension of the DFT+$U$ approach that includes on-site $U$ and inter-site $V$ electronic interactions. In this extended Hubbard functional, the Hubbard parameter $V$ represents the strength of the Coulomb interactions between electrons on neighboring sites. The aim of this formalism is to improve the accuracy of the DFT+$U$ scheme for materials where hybridization between orbitals belonging to different atoms is important. When performing DFT+$U$+$V$ calculations, the predicted properties will strongly depend on the $U$ and $V$ values. In the past Hubbard parameters were often  determined empirically by fitting to experimental properties. However, not only is the choice of empirical Hubbard parameters not unique, as fitting to different experimental properties results in different values, but it was shown that fitting to band gaps or structural parameters will not generally result in defect energetics that agree with experiments~\cite{Wang2006, Lutfalla2011, Getsoian2013, Capdevila-Cortada2016}. Linear-response theory (based on supercell calculations) was used  to derive the Hubbard parameters in different materials from first-principles, thus removing the ambiguity associated with the empirical determination of these interdependent parameters and establishing the DFT+$U$+$V$ method as an accurate and versatile approach for materials with vastly different properties~\cite{Campo2010}. Recently, Timrov \textit{et al.}~\cite{Timrov2018} reformulated the linear-response calculation of Hubbard $U$ within density-functional perturbation theory (DFPT). This formulation is computationally cheaper due the use of sums over monochromatic (wave-vector-specific) perturbations in primitive cells, instead of finite-differences between supercell calculations. It also results in better numerical stability and convergence as well as a higher level of automation of the computational protocol. This formulation was recently extended to also yield the inter-site $V$ parameters~\cite{Timrov_inprep}. 

Defect formation in transition-metal oxides can induce local perturbations of the chemical environment of Hubbard sites around the defect that may not be properly described by a single global $U$ as done in conventional DFT+$U$. We therefore recently suggested a self-consistent site-dependent (SC-SD) DFT+$U$ approach, in which $U$ values are computed using DFPT for all inequivalent Hubbard sites around a defect, using a self-consistent procedure during which $U$ parameters and the geometry of the system are recomputed in an iterative fashion until convergence within given thresholds~\cite{Ricca2019}. $U$ values were found to depend on the distance of the Hubbard site from the defect, its coordination number, its oxidation state, and on the magnetic order of the host material. This site-dependence was found to strongly influence the properties related to the defect energetics, particularly for semiconducting and insulating materials, where filled localized defect states may form in the band gap. The same approach could be easily extended to DFT+$U$+$V$ calculations by computing self-consistent site-dependent $U$ and $V$ parameters for all inequivalent Hubbard sites and site pairs around a defect. It is worth to mention here that the same site-dependence would apply, in principle, also to the more frequently used but also computationally more expensive hybrid functionals~\cite{jaramillo2003, kaupp2006, kaupp2007, kaupp2007b, kaupp2007c, shimazaki2015theoretical}. Instead, in the most common hybrid functionals, the fraction of exact exchange, which affects the predicted defect energy levels and formation energies, is a global value, which may not be suitable to properly describe the excess charge localization in defective structures~\cite{Zhang2005hybrid, Franchini2007, onishi2008hybrid}.

In this work, we apply the DFT+$U$+$V$ approach to study V$_\mathrm{O}$ in STO using self-consistent $U$ and $V$ values computed using DFPT. Inter-site interactions between Ti-$3d$ and O-$2p$ states are included to account for the previously reported mixed ionic-covalent character of the Ti--O bond. The effect of site-dependent $U$ and $V$ is also addressed. Due to the puzzling and contradictory conclusions of previous theoretical works and due to the difficulties in performing a direct comparison with the widely scattered and often also contradictory experimental data, our goal is not to benchmark the DFT+$U$+$V$ method with respect to the available theoretical and experimental data for V$_\mathrm{O}$ in STO, but to rationalize these results using a method (DFT+$U$+$V$) that is accurate and cost-effective at the same time. For this reason, we perform a systematic study of oxygen-deficient STO, in which cell size, STO crystalline phase, and spin polarisation are taken into account. Results are compared with data obtained, within the same computational framework, from standard DFT, DFT+$U$ (with SC and SC-SD $U$ values), and in selected cases also from hybrid functionals. Given the large quantity of data, we report in the paper only selected results that best illustrate the main trends and conclusions~\footnote{Stoichiometric data is discussed based on the 40-atom cell, for which we also performed hybrid functional calculations. For defective structures, the density of states are always shown for the 80-atom cell as smaller cell sizes do not allow a clear identify the defect band due its dispersion, while larger cells result in a defect band merged with the conduction band. Self-consistent site-dependent Hubbard parameters are  discussed for the 40-atom cell, where the effect of the long-range dependence of these parameters on the formation energy can be more clearly rationalized due to the smaller number of Ti sites or Ti-O pairs. Finally, we mainly report results for the low temperature AFD phase, except for the discussion of the magnetic properties of the neutral oxygen vacancy, where results for the cubic phase are shown for direct comparison with previously published results.}. The whole set of results is available on Materials Cloud~\footnote{DOI: 10.24435/materialscloud:2020.0011/v1}. 

\section{\label{sec:review}Oxygen Vacancies in STO}

Despite the large amount of both experimental and theoretical work dedicated to characterizing oxygen-deficient STO, the nature of oxygen vacancies in this material still remains debated. On one hand, one has to consider the complex experimental measurements necessary for the characterization of these defects and the varied nature of the analyzed samples, especially in terms of type (bulk or surfaces) and defect concentration, which could explain the diversity of the experimental results reported in the literature. On the other hand, theoretical calculations based on DFT often result in contradictory results depending on the exchange and correlation functional, the considered magnetic properties of the defective system, the size of the supercell used to simulate STO, as well as the STO phase. In the following we will shortly review the main findings reported in  literature.

 An insulator-to-metal transition taking place at extremely dilute electron doping (on the order of 10$^{19}$~cm$^{-3}$) and superconducting behavior with a transition temperature below 1 K were experimentally observed in oxygen-deficient STO~\cite{Calvani1993, Schooley1964, Koonce1967, PHILLIPS1969356, Binnig1980, Ohtomo2007}. However, Hall measurements suggest a mobile carrier density lower than the expected two carriers per V$_\mathrm{O}$, especially for higher defect concentration~\cite{GONG1991320}. Transport measurements indicate that the formation of oxygen vacancies is associated with the appearance of shallow donor levels~\cite{Tufte1967, Ohtomo2007}, but optical emission and absorption spectra of oxygen-deficient STO show peaks due to localized defect states in the band gap~\cite{Kan2005, Shibuya2007, Zhang2008, Kim2009}, especially for STO surfaces~\cite{Henrich1978, Courths1980, ADACHI1999272}. For STO single crystals with low carrier density, the V$_\mathrm{O}$ ionization energy was found to range from 0.07 to 0.16 eV depending on the carrier density and the degree of compensation by residual acceptors~\cite{Tufte1967, Lee1971}. Moos and H\"{a}rdtl~\cite{Moos1997} on the other hand find the redox level for the ionization from the neutral V$_{\mathrm{O}}^{\bullet \bullet}$ to the singly charged V$_{\mathrm{O}}^{\bullet}$ to lie 3 meV below the conduction band (CB) and the redox level for the ionization from the doubly positively charged V$_{\mathrm{O}}^X$ to the singly charged V$_{\mathrm{O}}^{\bullet}$ to lie deeper at about 0.3 eV from the CB. From their data it is also possible to extrapolate the V$_\mathrm{O}$ formation enthalpy at 0 K to 6.1 eV~\cite{Evarestov2006, Evarestov2012}, which is in line with previous reports~\cite{Schwarz01051975}.
 
\begin{figure}
 \centering
 \includegraphics[width=0.95\columnwidth]{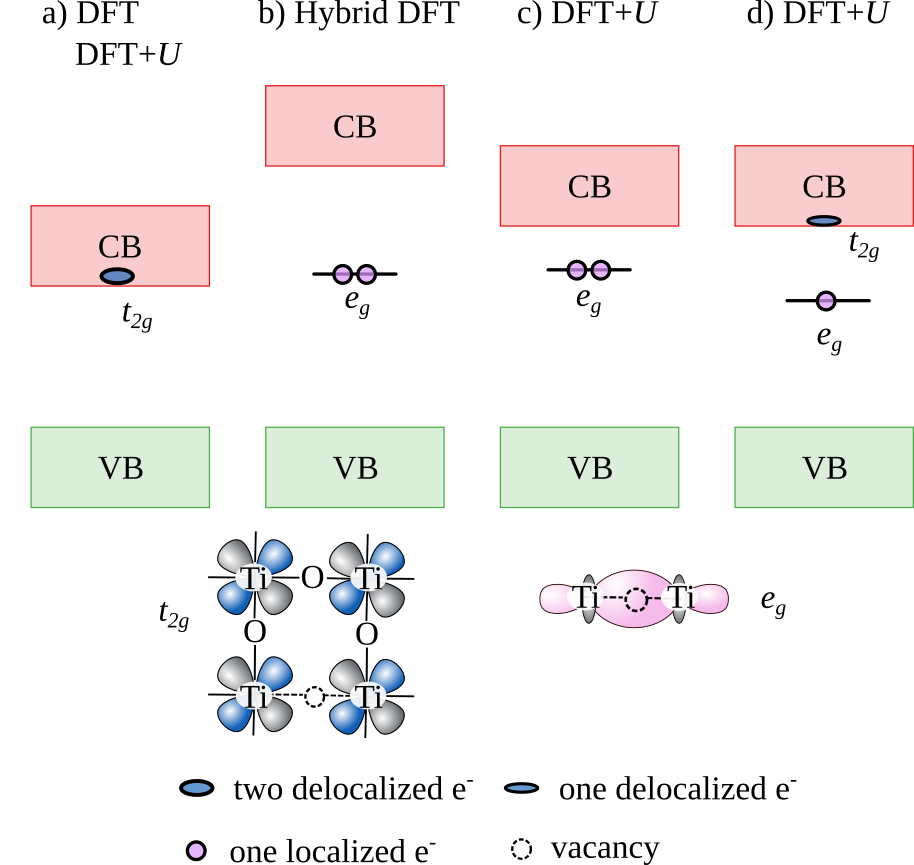}
 \caption{Schematic representation of the different localization schemes reported in theoretical studies of oxygen-deficient STO where a) standard DFT or DFT+$U$~\cite{Tanaka2003, Luo2004, Cuong2007}, b) DFT with hybrids~\cite{Evarestov2006, Ricci2003, Alexandrov2009, Zhukovskii2009, ZHUKOVSKII20091359,Mitra2012}, c) DFT+$U$~\cite{Cuong2007, Hou2010, Lin2012, Mitra2012, Choi2013}, and d) DFT+$U$ for the triplet state were used~\cite{Hou2010,Choi2013}. CB and VB represent the conduction and valence band, respectively. The position of the CB schematically represents the effect of the different DFT methods on the predicted STO band gap.}
\label{fig:localization_schemes}
\end{figure}

Theoretical studies based on DFT did not result in a coherent picture that explains all experiments. Indeed, several contradictory results for the electronic structure of V$_\mathrm{O}$ defects in STO were reported, depending on the exchange-correlation functional, the supercell size, the cubic or AFD phases of STO as well as the treatment of spin polarization and atomic relaxations. Several localization schemes for the two electrons associated with an isolated neutral V$_\mathrm{O}$ were suggested: i)~a delocalized state with 3$d$-$t_{2g}$ character in the CB (see Fig.~\ref{fig:localization_schemes}a), ii)~a deep doubly-occupied state with 3$d_{z^2-r^2}$ character in the band gap localized on the two Ti atoms adjacent to the defect (see Fig.~\ref{fig:localization_schemes}b), iii)-a shallow doubly-occupied 3$d_{z^2-r^2}$ state close to the CB (see Fig.~\ref{fig:localization_schemes}c), and iv)~a singly-occupied 3$d_{z^2-r^2}$ state lying below the CB coupled with one electron in delocalized 3$d$-$t_{2g}$ CB states (see Fig.~\ref{fig:localization_schemes}d)~\cite{Hou2010}. Schemes i)~and iv)~with electrons in the CB could explain the temperature independence of the free carrier density and the low temperature conductivity~\cite{Ohtomo2007, GONG1991320, Moos1997}. However, scheme i)~ cannot explain the reported optical properties~\cite{Kan2005, Shibuya2007, Zhang2008, Kim2009}, which are compatible with the presence of in-gap defect states. It is worth to note here that deep defect states were often reported to stem from surface Ti$^{3+}$~\cite{Henrich1978, Courths1980, ADACHI1999272} and that the interpretation of these measurements is not unambiguous since in-gap levels in single crystals could be attributed to defects or defect clusters other than isolated V$_\mathrm{O}$~\cite{Shanthi1998, djermouni20109, Choi2013, Jeschke2015, Cuong2007}. Hall measurements~\cite{GONG1991320} and the ionization energies for the redox level associated with V$_\mathrm{O}$~\cite{Moos1997} seem to support scheme iv) with one electron localized in an in-gap state and one delocalized in the CB. This implies, however, the presence of singly charged oxygen vacancies in a large temperature range, but paramagnetic V$_{\mathrm{O}}^{\bullet}$ were not observed by electron paramagnetic resonance (EPR) spectroscopy~\cite{Longo2008} at low temperatures.

The main source of these discrepancies in theoretical studies stems from the flavor of the DFT functional. Standard LDA and GGA functionals suggest that V$_\mathrm{O}$ induce a delocalized state in the CB~\cite{Astala2001, Astala200181, Tanaka2003, Luo2004, Cuong2007} (see also Fig.~\ref{fig:localization_schemes}a). The strong underestimation of the STO band gap by almost half of the experimental value of 3.25 eV~\cite{vanBenthem2001} can, however, result in the defect state to lie in the CB rather than in the band gap (cf.  Figs.~\ref{fig:localization_schemes}a and~\ref{fig:localization_schemes}b). Several attempts to correct the STO band gap using DFT+$U$ were performed, but results still suggest several scenarios for the position of the defect level~\cite{Cuong2007, Hou2010, Lin2012, Mitra2012, Choi2013}. We should note here that due to the  formally empty Ti-$d$ states the physical motivation for the DFT+$U$ correction is not clear for STO and that the choice of Hubbard $U$ strongly affects the results, thus not justifying the use of empirical $U$ values. In the majority of studies, an effective $U$ value of 4.36~eV~\cite{Cuong2007, Kim2009, Hou2010, Choi2013} is applied, which however results in a band gap still significantly smaller (by about 1~eV) than experiment, thus resulting either in shallow (about 0.1-0.4~eV below the CB)~\cite{Cuong2007, Choi2013, Hou2010} or delocalized~\cite{Kim2009, Hou2010, Choi2013} defect states (cf. Figs.~\ref{fig:localization_schemes}a,~\ref{fig:localization_schemes}c and~\ref{fig:localization_schemes}d)). Discrepancies in results obtained with similar $U$ are most likely due to structural differences, different spin polarization treatment or different computational details, as we will discuss in Sec.~\ref{sec:results}.

Larger but unphysical (for Ti atoms with $d^0$ configuration) $U$ values of about 8 eV~allow to reproduce the experimental band gap and yield deeper defect levels similar to those obtained with hybrid functionals~\cite{Lin2012, Mitra2012}. With these methods V$_\mathrm{O}$ defects were found to always be associated with deep in-gap states 0.7 to 1.2~eV below the CB that are derived from the 3$d_{z^2-r^2}$ orbitals of the two Ti adjacent to the vacancy~\cite{Evarestov2006, Ricci2003, Alexandrov2009, Carrasco2005, Zhukovskii2009, ZHUKOVSKII20091359, Mitra2012}. Lin \textit{et al.}~\cite{Lin2012} suggested that these in-gap states with $e_g$ character are observed within these approaches due to a smaller $t_{2g}-e_g$ crystal field splitting for stoichiometric STO. Within standard DFT, this splitting is large and the defect state is located in the CB $t_{2g}$ bands, while a smaller crystal field splitting, as the one provided by hybrid DFT, allows the formation of a localized $e_g$-derived state in the gap. This localization of electrons on Ti sites adjacent to the defect was often justified with Ti$^{3+}$ found in photoemission studies on defective STO surfaces, which could however behave differently from bulk sites due to the different coordination. This localization also corresponds to very high V$_\mathrm{O}$ formation energies of 7-10~eV. It was suggested~\cite{Mitra2012} that despite hybrid DFT results being similar those obtained with a large Hubbard $U$ correction on the Ti-3$d$ states, the origin of the in-gap state is different, since with DFT+$U$ the unoccupied bands are pushed up, while with hybrid functionals, the occupied defect state is also pushed down in energy. It was also suggested~\cite{Hou2010} that the very deep position of the defect state, which contradicts conductivity measurements, is due the overestimation of the STO band gap (by about 0.3~eV) as a consequence of the linear combination of atomic orbitals (LCAO) basis set used in the majority of the hybrid functional calculations~\cite{Evarestov2006, Ricci2003, Alexandrov2009, Zhukovskii2009, ZHUKOVSKII20091359}. It was, however, shown that band gaps are similar for plane-wave and LCAO calculations in the same supercell, implying that the overestimation is due to the hybrid functional~\cite{Mitra2012}.

In several studies it was observed that, apart from the DFT functional, the vacancy formation energy and the position of the defect state are highly sensitive to the size of the supercell. Buban \textit{et al.}~\cite{Buban2004} using LDA found that the defect state changes from deep to shallow going from a 40- to a 160-atom (or the even larger 320-atom) supercell, with mainly 3$d_{z^2-r^2}$ character for the 40-atom case and 3$d$-$t_{2g}$ character for cells larger than 160-atoms. Similar results were obtained using DFT+$U$ by Choi \textit{et al.}~\cite{Choi2013}. Evarestov~\cite{Evarestov2006} observed an analogous behavior with the hybrid functional B3PW: not only does the reduced interaction between periodic images of the V$_\mathrm{O}$ result in a strongly reduced dispersion of the defect state for larger cells (0.15 eV in the 80- and 0.02~eV in the 320-atom cell) but its position moves from 0.69 to 0.49~eV below the CB when increasing the supercell size from 80 to 320 atoms. These changes are reflected in the computed formation energies, which change by about 0.5-1.3~eV between the 40- and 320-atom supercell~\cite{Carrasco2005, Buban2004, Evarestov2006}. The size of the supercell was also found to affect the calculated formation energy: using the B3PW hybrid functional an orthorhombic 80-atom cell results in a formation energy of 7.73 eV while a cubic 135-atom was found to yield a formation energy about 0.2~eV larger despite the bigger cell size~\cite{Evarestov2006}. Finally, also the STO phase can affect the results. Choi \textit{et al.}~\cite{Choi2013} using a 1080-atom supercell reported that in cubic STO there are no thermodynamic transition levels in the band gap, while for the AFD phase they computed the V$_{\mathrm{O}}^{\bullet \bullet}$/V$_{\mathrm{O}}^{\bullet}$ transition level at about 0.1 ~eV below the CB minimum. It was suggested that the drop in the AFD to cubic transition temperature from 105~K for stoichiometric to about 98~K for oxygen-deficient STO~\citep{Hunnefeld2002} could be explained considering that AFD-like oxygen-octahedron rotations are induced in the vicinity of a V$_\mathrm{O}$ in cubic STO~\cite{Buban2004, Choi2013}.

Structural relaxations were indeed shown to be important, especially for the existence of localized states but contradictory results were reported also in this case. Evarestov \textit{at al.}~\cite{Evarestov2006} reported that the V$_{\mathrm{O}}^{\bullet \bullet}$ formation energy is reduced by 1.5-2.0~eV when the positions of all atoms in the supercell are optimized, while Alexandrov \textit{et al.}~\cite{Alexandrov2009} reported only small relaxation energies of about 0.1~eV. These discrepancies were explained with the different DFT method and basis set, the first results being obtained with standard DFT and plane-waves and the second using hybrid functionals with LCAO. While in the first case the two electrons due to the defect are delocalized in the CB, resulting in a doubly positively charged V$_\mathrm{O}$ with strong repulsive interactions with the surrounding Ti cations, in the second case, the hybrid functional yields an F-center with the electrons localized in the vacancy. Other works~\cite{Carrasco2005, Evarestov2012} using a similar approach to Alexandrov \textit{et al.}~\cite{Alexandrov2009} report large relaxations for the first and second nearest neighbor atoms around the defect and localization of the extra electrons on the Ti adjacent to the defect. Not only the magnitude but also the direction of relaxations varies between different reports. Mitra \textit{et al.}~\cite{Mitra2012} analyzed the displacements for different functionals within a similar computational setup and linked them to the observed electronic structure. Within LDA, the Ti adjacent to the vacancy move away from each other, while with the hybrid functional HSE06~\cite{heyd2006j} they approach each other. Using HSE06, a localized in-gap state exists even before structure relaxation, while this is not the case with LDA. This implies that electrons in a localized state lead to approaching Ti atoms, while delocalized CB electrons have the opposite effect.

A final source for discrepancy is the treatment of the magnetic structure of the defect. Early (semi)local DFT~\cite{Buban2004, Astala2001, ZHANG20121770} and hybrid functional~\cite{Alexandrov2009, Ricci2003} calculations reported the diamagnetic closed shell (singlet) state for the neutral oxygen vacancy to be more stable than the spin-polarized open shell (triplet) solution. This is despite the fact that standard DFT calculations predict shallow or delocalized defect states while hybrid functionals result in deep in-gap states. DFT+$U$~\cite{Cuong2007, Kim2009} results without spin polarization were often found to provide the same results as uncorrected (semi)local DFT. In more recent DFT+$U$ studies~\cite{Hou2010, Choi2013}, different possible magnetic solutions were investigated by fixing the distance between Ti atoms adjacent to the neutral V$_\mathrm{O}$ at a series of values and optimizing the other atomic positions in different magnetic states. For the non-polarized (singlet) state, two ground states were found, corresponding either to a delocalized solution or to a shallow defect state, again in line with previous reports. For the spin-polarized (triplet) solution, a ground state, characterized by one electron localized in the band gap and one delocalized in the CB was found, which is more stable compared to all of the non-polarized solutions. Taking the magnetic properties of the V$_\mathrm{O}$ defect into account is thus fundamental in understanding the origin of the discrepancies between theoretical results, even when obtained with the same DFT flavor, as in the case of DFT+$U$ results.

\section{Methods}
\label{sec:compdetails}

All calculations were performed with {\sc{Quantum ESPRESSO}}~\cite{giannozzi2009quantum,Giannozzi2017} using the PBEsol~\cite{perdew2008pbesol} exchange-correlation functional and ultrasoft pseudopotentials~\cite{vanderbilt1990soft} with Sr($4s$,$4p$,$5s$), Ti($3s$,$3p$,$4s$,$3d$), and O($2s$,$2p$) valence states \footnote{Ultrasoft pseudopotentials from the PSLibrary version 1.0.0. (\url{https://dalcorso.github.io/pslibrary/}): Sr.pbesol-spn-rrkjus\textunderscore psl.1.0.0.UPF, Ti.pbesol-spn-rrkjus\textunderscore psl.1.0.0.UPF, and O.pbesol-n-rrkjus\textunderscore psl.1.0.0.UPF.}~\cite{DALCORSO2014337}. Wavefunctions were expanded in plane waves with a cutoff of 40 Ry for the kinetic energy and 320~Ry for the charge density. Gaussian smearing with a broadening of 0.01 Ry was used in all calculations, including plotting the density of states (DOS).

STO structures were described using different supercell sizes: $2\times2\times2$, $2\sqrt{2} \times2\sqrt{2} \times2$ , $3\times3\times3$, and $4\times4\times4$ supercells of the 5-atom primitive cell containing 40, 80, 135 and 320 atoms, respectively. The $3\times3\times3$ cell was used only for the cubic phase, since the octahedral rotations of the AFD phase are incommensurate with this cell. Monkhorst-Pack~\cite{monkhorst1976special} \textbf{k}-point meshes of size $4\times4\times4$, $3\times3\times4$, $3\times3\times3$ were applied for the 40- 80- and 135-atom cell, respectively, while only $\Gamma$ point sampling was performed for the largest 320-atom cell. Grids were doubled along every dimension for plotting the DOS.

V$_\mathrm{O}$ defects were created by removing one oxygen atom from each of the considered supercells. Neutral (V$_\textrm{O}^{\bullet \bullet}$), singly (V$_\textrm{O}^{\bullet}$), and doubly (V$_\textrm{O}^\textrm{X}$) positively charged V$_\mathrm{O}$ were created by adjusting the number of electrons. For charged defects, calculations were performed in presence of a jellium background, necessary to avoid divergence of the electrostatic potential. For stoichiometric STO calculations, both ionic positions and cell parameters were relaxed during geometry optimization, while, for defective STO, only atomic positions were optimized while keeping the lattice vectors fixed at the optimized values of the stoichiometric bulk in order to mimic the dilute defect limit. In all cases, atomic forces were converged to within $5.0\cdot 10^{-2}$ eV/\AA, while energies were converged to within $1.4\cdot 10^{-5}$ eV.

DFT+$U$ calculations were performed within the rotationally-invariant formulation by Dudarev \textit{et al.}~\cite{dudarev1998electron} by applying a Hubbard $U$ correction~\cite{anisimov1991band, anisimov1997first} on Ti 3$d$ states with self-consistent $U_\mathrm{SC}$ and self-consistent site-dependent $U_\mathrm{SC-SD}$ values computed via DFPT~\cite{Timrov2018}. For DFT+$U$+$V$ calculations, we additionally applied the Hubbard $V$ correction to the inter-site interaction between the Ti-3$d$ and O-2$p$ states using $V$ values computed self-consistently (and site-dependently) within the DFPT approach~\cite{Timrov2018}. The method for the calculation of these parameters has been introduced in Ref.~\cite{Ricca2019}. Here, we shortly recall that $U$ and $V$ values are obtained through an iterative procedure that involves perturbing all inequivalent Hubbard sites in the structure and in which both the ionic and electronic structure are corrected using updated $U$ and $V$ values until convergence.

In the stoichiometric system all Ti sites are crystallographically and chemically equivalent and can thus be described by a global $U$ value ($U_\mathrm{SC}$) computed self-consistently by perturbing a single Ti. For the determination of the $V$ parameters, one needs to consider that due to octahedral rotations O sites can be inequivalent already in the stoichiometric system, implying that more than one O atom needs to be perturbed. We however observed that differences in $V$ values are so small that the Ti--O interaction in the stoichiometric material can also be described by a global $V_\mathrm{SC}$ value. In order to simplify and automate these calculations, atoms were selected to be perturbed if their unperturbed atomic occupations differed by more than 10$^{-6}$. For defective supercells, Hubbard parameters were computed by perturbing all inequivalent sites created upon defect formation (see Sec.~\ref{sec:methods_scsd} in the supporting information (SI) \footnote{See supplemental material at [LINK], for additional structural and electronic data for stoichiometric and oxygen deficient \ce{SrTiO3}.} for more details). DFPT calculations were performed with a $2\times2\times2$ mesh to sample \textbf{q} space~\cite{Timrov2018} in the 40-atom cell, while sampling was restricted to the $\Gamma$ point for larger cells. A convergence threshold of 0.01 eV was applied for the self-consistence of both $U$ and $V$ values. The self-consistent field (SCF) calculations preceding DFPT calculations were performed with atomic orbitals orthogonalized using L\"owdin's method~\cite{Lowdin1950} for the Hubbard manifold, while structural optimizations were performed without orthogonalization. This is necessary due to technical difficulties in implementing forces and stresses for DFT+$U$ and DFT+$U$+$V$ with orthogonal basis sets but is expected to only lead to marginal errors~\cite{Cococcioni2018}.

Hybrid DFT SCF calculations for the stoichiometric and defective 40-atom STO cells were performed with the HSE functional~\cite{heyd2003hybrid,heyd2006j}, where a percentage ($0\le\alpha\le 1$) of exact exchange is mixed with the PBE~\cite{perdew1996generalized} functional. To examine the effect of the percentage of exact exchange on the defect properties, calculations were performed for $\alpha$ ranging from 0 to 0.25, this upper limit being the default for HSE06, while keeping the screening parameter at the default value for HSE (0.2 \angstrom$^{-1}$). 

The formation energy of an oxygen vacancy ($\textrm{V}_\textrm{O}$) in a charge state $q$ ($E_{f,\textrm{V}_\textrm{O}^q}$) was computed as described in Ref.~\onlinecite{freysoldt2014first}:
\begin{align}
E_{f,\textrm{V}_\textrm{O}^q} &= E_{\textrm{tot},\textrm{V}_\textrm{O}^q}-E_{\textrm{tot,stoic}}+\mu_\mathrm{O} \nonumber \\
&+ q \, [E_\textrm{V} + E_\textrm{F}] + E_{\textrm{corr}} \,,
\label{eq:formenerg}
\end{align}
where $E_{\textrm{tot},\textrm{V}_\textrm{O}^q}$ and $E_{\textrm{tot,stoic}}$ are the total energies of the defective and stoichiometric systems, $E_\textrm{F}$ ($0\le E_\textrm{F}\le E_\textrm{g}$) is the Fermi energy relative to the valence band maximum ($E_\textrm{V}$) of the stoichiometric system ($E_\textrm{g}$ being its band gap) and $\mu_\mathrm{O} = \frac{1}{2}\mu(\textrm{O}_2)+ \Delta \mu(\textrm{O})$ is the oxygen chemical potential with $\mu(\textrm{O}_2)$ obtained as the total energy of an O$_2$ molecule. We will show results in the oxygen-rich limit, \textit{i.e.} with $\Delta\mu(\textrm{O})=0$. For charged vacancies, a potential alignment term ($E_{\textrm{corr}}$) was also computed in order to realign the electrostatic potential of the defective supercell with the one of the stoichiometric system. This was done by calculating the difference in average electrostatic potential between the stoichiometric system and the charged defective one computed via spherically averaged electrostatic potentials at sites far from the defect~\cite{lany2008}. Finally, the thermodynamic transition level $\varepsilon (q_1/q_2)$ for two V$_{\textrm{O}}$ defects with charge states $q_1$ and $q_2$ was computed as:
\begin{multline}
\varepsilon (q_1/q_2) = \frac{E_{f,\textrm{V}_\textrm{O}^{q_1}}(E_\textrm{F} = 0) - E_{f,\textrm{V}_\textrm{O}^{q_2}}(E_\textrm{F} = 0)}{q_2 - q_1}\,.
\label{eq:translev}
\end{multline}
%

\section{\label{sec:results}Results and Discussion}

\subsection{Stoichiometric STO}\label{sec:results_stoich}

\begin{table}
\caption{Comparison of the calculated and experimental structural properties (lattice parameters \textit{a} in \AA, \textit{c}/\textit{a} ratio, and octahedral rotation angle around the $c$-axis $\theta$ in degrees) for the AFD and cubic phases obtained in a 40-atom cell.}
\begin{tabular*}{\columnwidth}{@{\extracolsep{\fill}}llcclll}
\hline
\hline
Phase               & Method      							& $U$  & $V$    & $a$ & $c$/$a$& $\theta $ \\
\hline
\multirow{4}{*}{AFD} & GGA         							& - 		& - 		& 3.860 	& 1.006  	 & 5.69   \\
                     & GGA+$U_\mathrm{SC}$     				& 4.48 		& - 		& 3.857 	& 1.012  	 & 7.81   \\
                     & GGA+$U_\mathrm{SC}$+$V_\mathrm{SC}$ 	& 5.34 		& 1.27 		& 3.841 	& 1.009  	 & 6.47   \\
                	 & Exp.           						& -    		& -    		& 3.898$^a$ & 1.006$^a$ & 2.1$^b$   \\
                	 \hline
\multirow{4}{*}{Cubic}   & GGA         							& -    & - & 3.870 	  & -      & -      \\
                     & GGA+$U_\mathrm{SC}$    				& 4.45 & - & 3.877 	  & -      & -      \\
                     & GGA+$U_\mathrm{SC}$+$V_\mathrm{SC}$ 	& 5.35 & 1.31 & 3.855 	  & -      & -      \\
                 	 & Exp.           						& -    & -    & 3.900$^a$ & -      & -     \\
\hline
\hline
\end{tabular*}
\begin{flushleft}
\small
$^a$ Ref.~\cite{Cao2000} data at 65 and 105 K for the AFD and cubic phases of STO, respectively.\\
$^b$ Ref.~\cite{Unoki1967} data at 4.2 K.\\
\end{flushleft}
\label{tbl:bulkproperties}
\end {table}
We begin by comparing the structure and electronic properties obtained for stoichiometric STO at the GGA, GGA+$U_\mathrm{SC}$, and GGA+$U_\mathrm{SC}$+$V_\mathrm{SC}$ levels of theory. We will discuss the results for the smallest 40-atom cell for which we performed also more expensive HSE06 calculations. Similar conclusions can, however, also be derived for larger cell sizes as can be seen in Table~\ref{tbl:SI_sto_structure}. Table~\ref{tbl:bulkproperties} shows the respective structural properties together with the computed self-consistent values of the Hubbard parameters for both the AFD and cubic phases of STO. $U_\mathrm{SC}$ and $V_\mathrm{SC}$ values are similar for the two phases, but seem to be consistently slightly larger for the cubic phase, which could be a consequence of the different crystal environment of the Hubbard sites without octahedral rotations. We also note that the Hubbard $U$ calculated for DFT+$U$ are about 0.9~eV smaller than the $U$ computed within the DFT+$U$+$V$ approach as a consequence of the necessity to perturb neighboring ligand states when evaluating the inter-site $V$ parameters, which results in the removal of theses states from the ``screening'' manifold~\cite{Cococcioni2018}. 

In the cubic phase, GGA+$U_\mathrm{SC}$ seems to provide the best description of the lattice parameter $a$, expanding it with respect to GGA. In the AFD case, instead, GGA provides the best agreement with experiments. In both cases, GGA+$U_\mathrm{SC}$+$V_\mathrm{SC}$ results in the smallest $a$ because the inter-site interactions encourage the occupation of hybridized states, shortening the bonds. We note, however, that experimental lattice parameters at 0 K should be smalller than the ones in Table ~\ref{tbl:bulkproperties}, which could reduce the error associated with GGA+$U_\mathrm{SC}$+$V_\mathrm{SC}$. Moreover, the underestimation of $a$ in GGA+$U_\mathrm{SC}$+$V_\mathrm{SC}$ is also a consequence of the underlying DFT functional. Table~ \ref{tbl:SI_structure_pbe} shows the lattice parameters computed for the cubic 5-atom unit cell of STO with either the PBEsol functional, similar to the data in Table~\ref{tbl:bulkproperties}, or the PBE functional. Compared to PBEsol, optimized for the description of solids, PBE always results in overestimated $a$ values, with PBE+$U_\mathrm{SC}$+$V_\mathrm{SC}$ providing again an improved description compared to PBE+$U_\mathrm{SC}$ and an accuracy similar to PBE. However, PBE is generally associated with larger errors than PBEsol and it also results in a larger underestimation of the band gap, which is a key property for the description of oxygen-deficient STO, as we will discuss in the following. We note here that discrepancies between PBEsol results in Tables~\ref{tbl:bulkproperties} and~\ref{tbl:SI_structure_pbe} are due to numerical differences between the calculations for the 5- and 40-atom cells. Finally, we observe that GGA, GGA+$U_\mathrm{SC}$ and GGA+$U_\mathrm{SC}$+$V_\mathrm{SC}$ all result in much larger octahedral rotation angles than experiment, GGA+$U_\mathrm{SC}$+$V_\mathrm{SC}$ yielding smaller values compared to GGA+$U_\mathrm{SC}$. We note that octahedral rotation angles are best predicted by hybrid functionals~\cite{Wahl2008, Evarestov2011phonon, ElMellouhi2013structural}, which is likely due to the effect of these functionals on the empty Sr states~\cite{Aschauer_2014}, which are not affected by our DFT+$U$ or DFT+$U$+$V$. Also compared to GGA+$U_\mathrm{SC}$, GGA+$U_\mathrm{SC}$+$V_\mathrm{SC}$ results in an improved description of the $c$/$a$ ratio, which is often considered as a measure for the quality of DFT results~\cite{Wahl2008}.

\begin{figure}
 \centering
 \includegraphics[width=0.9\columnwidth]{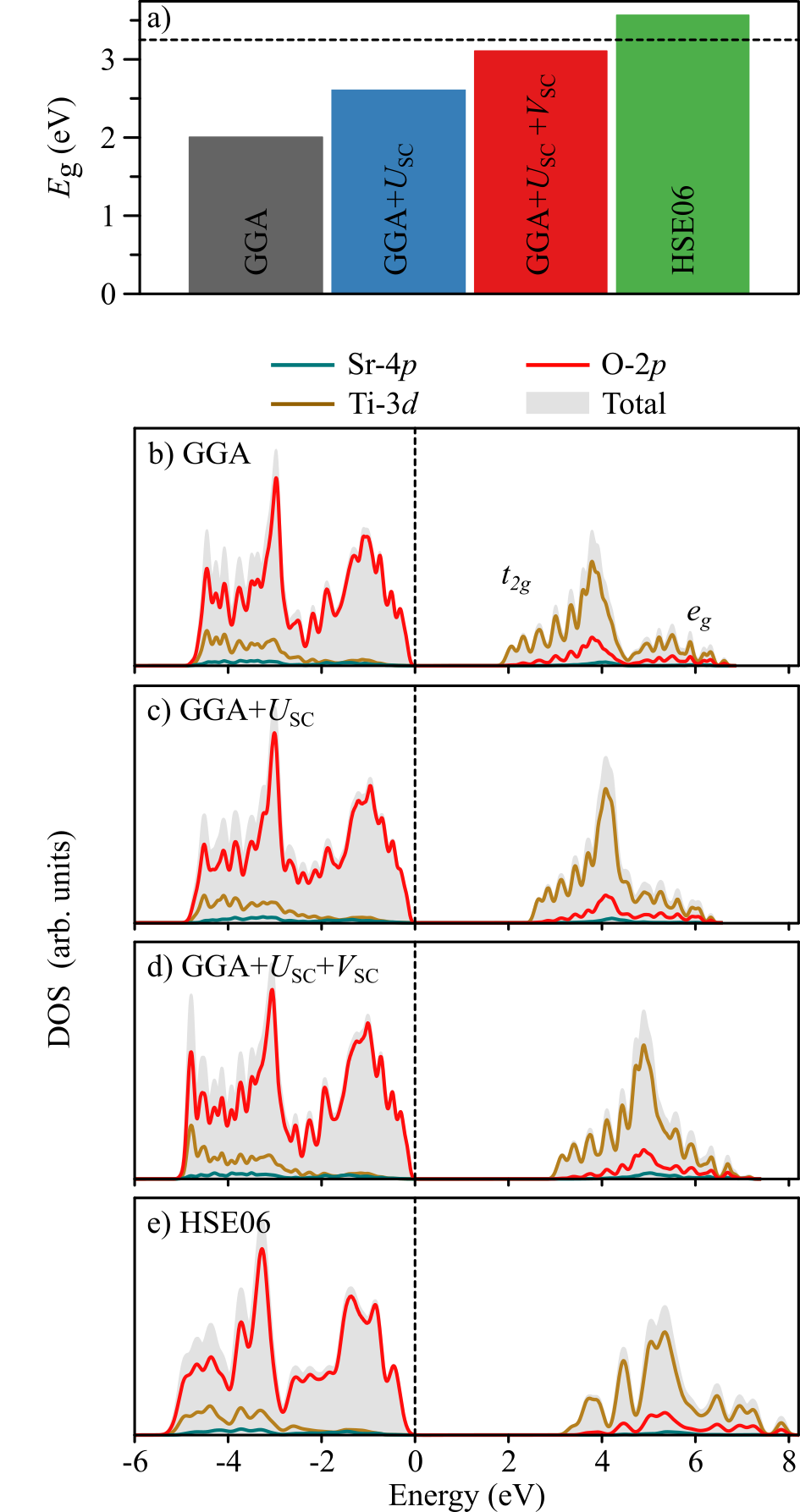}
 \caption{a) Band gap for the AFD phase of STO computed using a 40-atom supercell with different exchange-correlation functionals and b) total and projected density of states (DOS and PDOS, respectively) computed at different levels of theory. The zero of the energy scale was set at the top of the valence band in all cases. }
\label{fig:STO_bandgap_bulk}
\end{figure}
Fig.~\ref{fig:STO_bandgap_bulk}a shows a comparison of the band gap computed for the AFD phase of STO with different DFT functionals. The HSE06 result is obtained from of a SCF calculation on the GGA+$U_\mathrm{SC}$+$V_\mathrm{SC}$ structure. The band gaps of 2.00 and 2.62 eV obtained using GGA and GGA+$U_\mathrm{SC}$ are in line with previous works~\cite{Astala2001, Astala200181, Tanaka2003, Luo2004, Cuong2007, Choi2013, Hou2010} and are considerably lower than the experimental gap of 3.25 eV~\cite{vanBenthem2001}. The smallest error is associated with the value of 3.11 eV obtained with GGA+$U_\mathrm{SC}$+$V_\mathrm{SC}$, underestimating the band gap by only 0.14~eV with respect to experiments. HSE06, instead results in a band gap overestimated by about 0.32~eV, in agreement with previous reports~\cite{Carrasco2005, Buban2004, Evarestov2006}. Hence, GGA+$U_\mathrm{SC}$+$V_\mathrm{SC}$ provides the best agreement with experiments and at a much lower computational cost compared to hybrid functionals. While neglecting structural relaxations at the HSE06 level can affect these results, it is, however, also established that the percentage of exact exchange is a material-dependent quantity, which is generally determined by fitting to experimental data~\cite{ciofini2004, Franchini2007, Franchini2014}. In this sense, the Fig.~\ref{fig:STO_bandgap_hybrid} reports the band gap of STO as a function of the fraction of exact exchange, suggesting that a value of 20\% would result in a better agreement with experiment than the 25\% of exact exchange included in the standard formulation of this functional. The ambiguity associated with the fraction of exact exchange also affects the position of the defect level of oxygen vacancies in STO~\cite{Zhang2005hybrid, Franchini2014, onishi2008hybrid}, which will be discussed below in more detail. As opposed to HSE06 calculations, the DFT+$U_\mathrm{SC}$+$V_\mathrm{SC}$ approach we use here relies on Hubbard $U_\mathrm{SC}$ and $V_\mathrm{SC}$ parameters computed from first principles and with a self-consistent procedure that ensures the internal consistency of results. Hence, this approach does not rely on any empirical parameters and yields better results than hybrid functionals at a computational cost that is significantly lower. A similar trend can also be seen for the band gaps in the cubic phase (see Table~\ref{tbl:SI_sto_structure}), which are about 0.1-0.2 eV smaller than in the AFD phase, due the decrease in CB width by the octahedral rotations~\cite{ENG200394}.

Figures~\ref{fig:STO_bandgap_bulk}b-e show DOS and projected DOS (PDOS) of stoichiometric STO computed at different levels of theory: DFT, DFT+$U$, DFT+$U$+$V$, and DFT with HSE06. The valence band (VB) of STO is composed of O-2$p$ states with rather small contributions from Ti-3$d$ orbitals, while the conduction band is dominated by empty Ti-3$d$ states. In particular, the bottom of the CB is constituted mostly of $t_{2g}$ states with $e_g$ states lying at higher energies. At the GGA level in Fig.~\ref{fig:STO_bandgap_bulk}b, the $t_{2g}$ and $e_g$ bands can be clearly distinguished below and above 4.5~eV, respectively. When including the on-site Hubbard $U$ correction (see Fig.~\ref{fig:STO_bandgap_bulk}c), the empty Ti-3$d$ states are only slightly pushed to higher energies, in line with the nominal $d^0$ configuration of Ti in this material, and the separation between the $t_{2g}$ and $e_g$ states is also only slightly reduced. When both on-site $U$ and inter-site $V$ interactions are included, not only are the Ti states of the CB pushed to even higher energies, resulting in a larger band gap, but the $t_{2g}$-$e_g$ crystal field splitting is reduced (see Fig.~\ref{fig:STO_bandgap_bulk}d), which is in good agreement with the qualitative picture provided by HSE06 shown in Fig.~\ref{fig:STO_bandgap_bulk}e. As discussed in Sec.~\ref{sec:review}, the magnitude of the computed $t_{2g}$-$e_g$ crystal field splitting in stoichiometric STO has a strong influence on the predicted electronic structure of oxygen vacancies as we will discuss in Sec.~\ref{sec:V-O}.

\subsection{Oxygen-deficient STO}\label{sec:V-O}

We now proceed to the investigation of neutral as well as singly and doubly positively charged oxygen vacancies (V$_\textrm{O}^{\bullet \bullet}$, V$_\textrm{O}^{\bullet}$, and V$_\textrm{O}^{X}$, respectively, in Kr\"oger-Vink notation~\cite{KROGER1956307}). We will discuss the potential of using DFT+$U$+$V$ for the description of oxygen-deficient STO and how strongly the computed electronic properties and the formation energy for a neutral defect depend on the exchange-correlation functional, on the cell size, the crystal structure, on the treatment of the spin polarization, and on relaxation effects. We will mainly concentrate on the neutral defect since its properties are still widely debated. Finally, we will also show that the use of site-dependent Hubbard parameters should be carefully evaluated for band-like or shallow defect states.

\subsubsection{Electronic Properties}
\begin{figure}
 \centering
 \includegraphics[width=0.9\columnwidth]{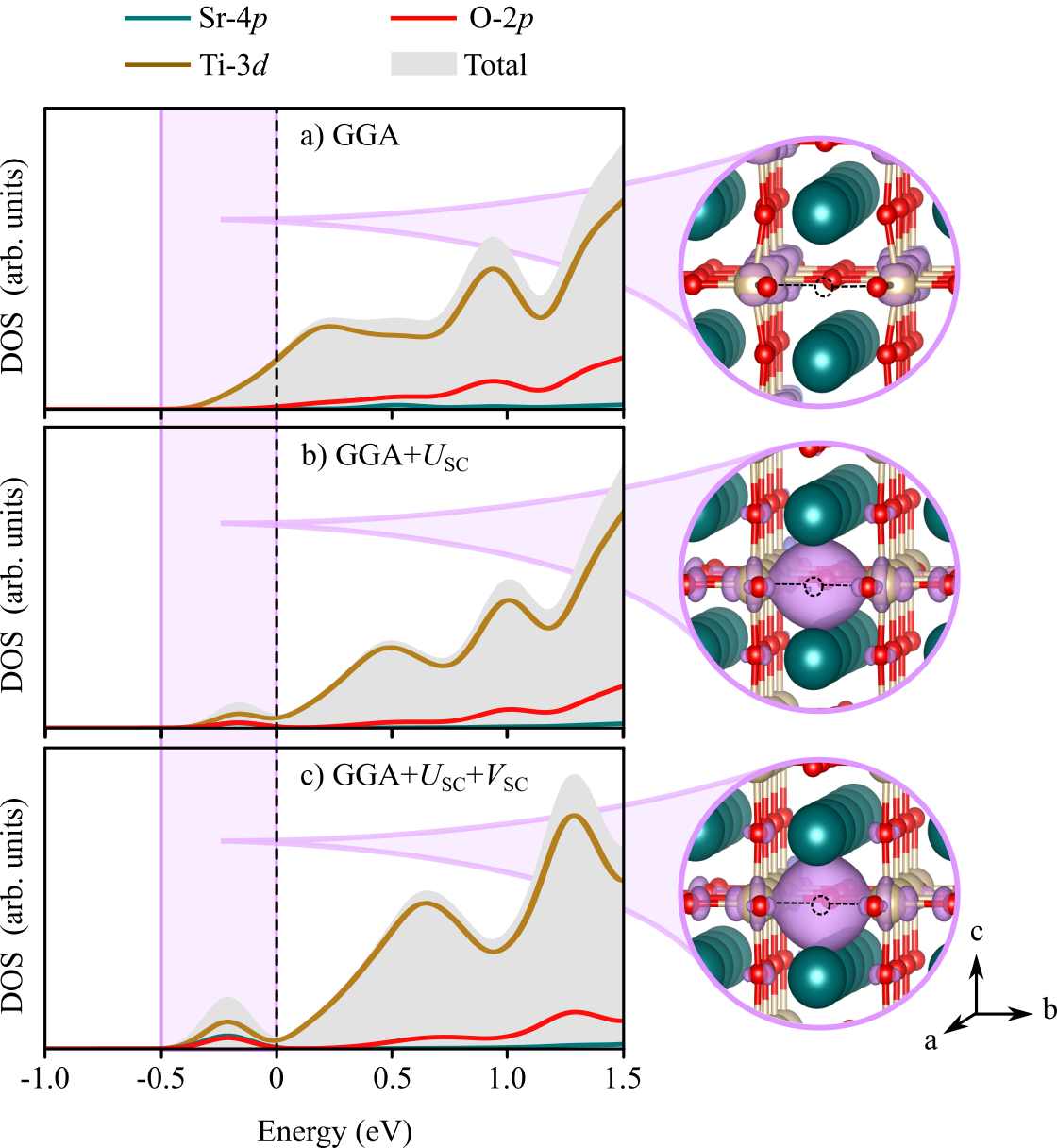}
 \caption{Projected density of states (PDOS) for a V$_\textrm{O}^{\bullet \bullet}$ in the 80-atom cell of the AFD phase of STO computed using a) GGA, b) GGA+$U_{\textrm{SC}}$, and c) GGA+$U_{\textrm{SC}}$+$V_{\textrm{SC}}$. The vertical dashed line indicates the position of the Fermi level. The isosurfaces ($10^{-2} \mathrm{e/\AA}^3$) on the right show the density in the energy range associated with the defect states (highlighted in purple in the PDOS).}
\label{fig:pdos_nVO}
\end{figure}
\begin{figure}
 \centering
 \includegraphics[width=0.9\columnwidth]{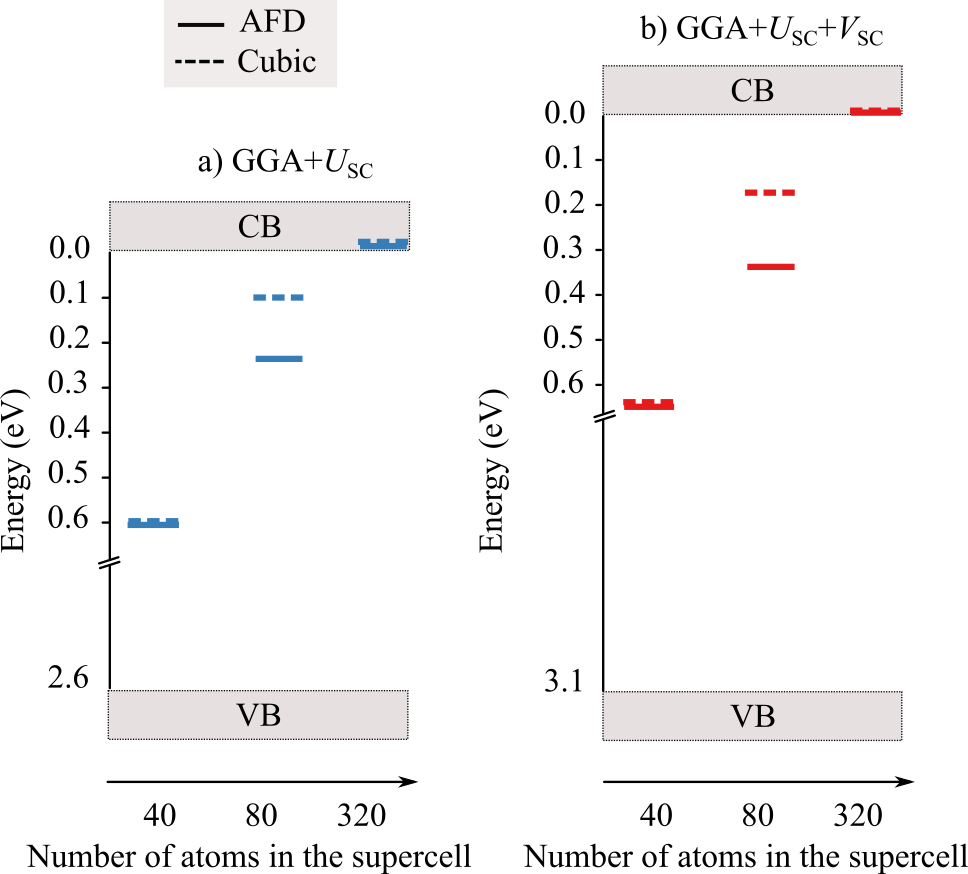}
 \caption{Energy level of the  electronic state of a V$_\textrm{O}^{\bullet \bullet}$ for different cell sizes obtained using a) GGA+$U_{\textrm{SC}}$ and b) GGA+$U_{\textrm{SC}}$+$V_{\textrm{SC}}$ in both the cubic (dashed lines) and the AFD phases of STO (solid lines). In both cases the zero is set to the CB minimum computed with the respective method.}
\label{fig:defect_pos_nVO}
\end{figure}

Figure~\ref{fig:pdos_nVO} shows the DOS computed for a V$_\textrm{O}^{\bullet \bullet}$ in the 80-atom cell of the AFD phase of STO with different methods. At the GGA level (see Fig.~\ref{fig:pdos_nVO}a) and in line with previous standard DFT calculations~\cite{Astala2001, Astala200181, Tanaka2003, Luo2004, Cuong2007}, the vacancy is associated with a delocalized defect state in the CB that has $t_{2g}$ character, as can be seen from the isosurface in Fig.~\ref{fig:pdos_nVO}a. At the GGA+$U_{\textrm{SC}}$ level, a F-center like defect state mainly localized on the vacancy site and formed by $e_{g}$ orbitals of the neighboring Ti atoms is observed instead (see Fig.~\ref{fig:pdos_nVO}b). However, this state is quite shallow and lies only 0.23~eV below the conduction band minimum (CBM) as shown in Fig.~\ref{fig:defect_pos_nVO}a and in Table~\ref{tbl:defectlevel}. A similar description is obtained when GGA+$U_{\textrm{SC}}$+$V_{\textrm{SC}}$ is used (see Fig.~\ref{fig:pdos_nVO}c), but the localization of the defect state is increased (about 0.35 eV below the CBM, see Fig.~\ref{fig:defect_pos_nVO}b and Table~\ref{tbl:defectlevel} for the 80-atom case). 

These results can be explained by the increase in the band gap (resulting in a lowering of the defect state with respect to the CB) and the reduction of the $t_{2g}$-$e_{g}$ crystal field splitting (facilitating the stabilization of an $e_g$ defect state) when going from standard DFT to DFT+$U$ and finally to DFT+$U$+$V$ as discussed in Sec.~\ref{sec:results_stoich}. Hence, DFT+$U$+$V$ provides results in better qualitative agreement with the description given by hybrid functionals (see Sec.~\ref{sec:review}) also for oxygen-deficient STO. From a quantitative point of view, hybrid functionals still provide much deeper defect states if results for the same cell size are considered: for the same 80-atom case we discussed above, Mitra \textit{et al.}~\cite{Mitra2012} reported a defect state lying 0.7~eV below the CBM (see Table~\ref{tbl:defectlevel}). Similarly, for the 40-atom cell we found the occupied V$_\textrm{O}^{\bullet \bullet}$ level to lie at 0.65 (see Fig.~\ref{fig:defect_pos_nVO}b and Table~\ref{tbl:defectlevel}) and about 0.9~eV (see Fig.~\ref{fig:pdos_hse}a and Table~\ref{tbl:defectlevel}) with GGA+$U_{\textrm{SC}}$+$V_{\textrm{SC}}$ and HSE06, respectively. This can be explained both by the fact that hybrid functionals push the defect state down in energy~\cite{Mitra2012} and by the overestimation of the band gap when 25\% of exact exchange is used in HSE06 (Sec.~\ref{sec:results_stoich}). If this fraction is reduced to 20\%, providing the best agreement with experimental and GGA+$U_{\textrm{SC}}$+$V_{\textrm{SC}}$ band gaps, the defect state moves towards the CBM (to about 0.75~eV, see Fig. \ref{fig:pdos_hse}a), thus reducing the disagreement between the two methods.

Unfortunately, the comparison of the above results with experiments is not straightforward, given the variety of the experimental data in literature (see Sec.~\ref{sec:review}), which is summarized in Table~\ref{tbl:defectlevel}. Ionization energies for STO single crystals derived from Hall or electrical conductivity measurements~\cite{Tufte1967,Lee1971, Moos1997, Raevski_1998} lie between 0.003 and 0.4~eV, suggesting that the formation of V$_\mathrm{O}^{\bullet \bullet}$ defects in STO crystals is associated with fairly shallow defect states, in line with the DFT+$U$+$V$ results. Similar results were obtained by photoluminescence \cite{Kan2005} and of ultra violet-visible (UV-vis) spectroscopy \cite{Ravichandran2011} of STO crystals and ultraviolet photoemission spectroscopy (UPS)~\cite{Henrich1978} of STO surfaces that locate the defect state at 0.4~eV and 0.1~eV from the CB, respectively. In other cases, much deeper defect states at about 1~eV were observed in UPS spectra~\cite{Henrich1978, Courths1980}. While these results were often used to validate hybrid functional results, they were obtained for STO surfaces and in some cases after ion (\ce{Ar+}) irradiation to induce defects. In summary, the experimental data in Table~\ref{tbl:defectlevel} seems to suggest that deeper localized states, as the ones predicted by hybrid functionals, are mainly observed for oxygen vacancies at STO surfaces rather than in the bulk, while results for STO single crystals are generally associated with fairly shallow defect states, in line with our DFT+$U$+$V$ results.

This improvement  in the description of the electronic properties of V$_{\textrm{O}}^{\bullet \bullet}$ is also reflected in the computed formation energies (see Fig.~\ref{fig:ef_nVO_cellsize}): while GGA and GGA+$U_{\textrm{SC}}$ result in $E_{f,\textrm{V}_{\textrm{O}}^{\bullet \bullet}}$ of about 5.5 eV for the 40-atom cell, the formation energy at the GGA+$U_{\textrm{SC}}$+$V_{\textrm{SC}}$ level (6.6 eV) is only slightly overestimated compared to the one obtained with HSE06 (6.3 eV) and to the experimental value extrapolated to 0 K (6.1 eV). As can be seen by comparing GGA+$U_{\textrm{SC}}$ and GGA+$U_{\textrm{SC}}$+$V_{\textrm{SC}}$ results (cf. Figs.~\ref{fig:defect_pos_nVO} and \ref{fig:ef_nVO_cellsize}), the formation energy increases for deeper (more localized) the defect states. This can be understood by the artificially small energetic cost associated with accommodating the two electrons in delocalized Ti states when the band gap is underestimated due to self-interaction errors. The same can be observed from the dependence of the defect localization, the formation energy, and STO band gap on the fraction of exact exchange in HSE (see Figs.~\ref{fig:STO_bandgap_hybrid} and \ref{fig:pdos_hse}).

Similar conclusions can be drawn regarding the effect of the Hubbard and extended Hubbard functionals on the electronic structure of the V$_\textrm{O}^{\bullet}$ defect, which are associated with a singly occupied defect state in the gap (see Fig.~\ref{fig:pdos_charged}a). This state is deeper compared to the doubly occupied one of V$_\textrm{O}^{\bullet \bullet}$ (cf. Fig.~\ref{fig:defect_pos_nVO} and Fig.~\ref{fig:defpos_scVO}) and its localization increases going from GGA+$U_{\textrm{SC}}$ to GGA+$U_{\textrm{SC}}$+$V_{\textrm{SC}}$. The effect of the cell size and STO phase is however different from the V$_\textrm{O}^{\bullet \bullet}$, as the V$_\textrm{O}^{\bullet}$ defect state becomes deeper with increasing cell size for the AFD phase with its larger band gap, while it becomes increasingly shallower for the cubic phase that has a smaller band gap. Finally, all methods provide a similar description of the electronic properties of V$_\textrm{O}^{X}$ where the empty defect state is merged with the CB (Fig.~\ref{fig:pdos_charged}b).

\subsubsection{Supercell size and STO phase}
\begin{figure}
 \centering
 \includegraphics[width=0.9\columnwidth]{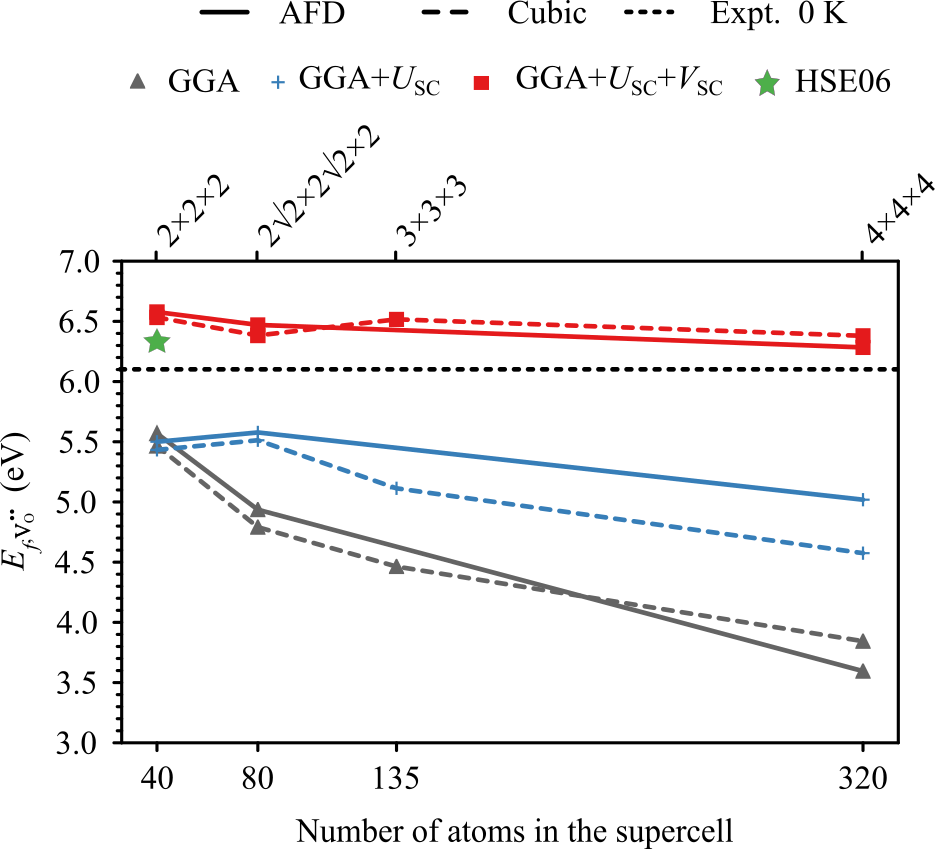}
 \caption{Formation energies computed for V$_\textrm{O}^{\bullet \bullet}$ with different methods and as a function of the cell size. The dotted black line indicates the experimental formation energy extrapolated to 0 K~\cite{Evarestov2006, Evarestov2012} and the green star indicates the value computed at the HSE06 level (computed for the AFD phase). Solid and dashed lines refer to data for the AFD and cubic phase, respectively.}
\label{fig:ef_nVO_cellsize}
\end{figure}
As highlighted in Sec.~\ref{sec:review}, the results obtained in previous theoretical reports may also differ because of different cell sizes. For this reason, we perform calculations with cell sizes ranging from 40- to 320-atoms for both the cubic and AFD phases of STO. As can be seen in Fig.~\ref{fig:defect_pos_nVO}, the electronic state associated with a V$_\textrm{O}^{\bullet \bullet}$ defect becomes shallower when going from the 40- to the 80-atom cell, while for even larger cell sizes the defect state is merged with the CB, independently of the functional and in line with observations by others using DFT, DFT+$U$ and hybrid functionals~\cite{Buban2004, Choi2013, Evarestov2006}. The V$_\textrm{O}^{\bullet \bullet}$ formation energy also strongly depends on the cell size when GGA or GGA+$U$ are used, which is a consequence of the incorrect description of the electronic properties of the defect state and the underestimation of the band gap within these two methods. The formation energy ($E_{f,\textrm{V}_\textrm{O}^{\bullet \bullet}}$) is indeed reduced by about 1.4-2.0 and 0.3-0.6~eV with GGA and GGA+$U_{\textrm{SC}}$, respectively, when going from the 40-atom to the 320-atom cell. Instead, GGA+$U_{\textrm{SC}}$+$V_{\textrm{SC}}$ results in formation energies that are fairly constant with cell size and in very good agreement with the experimental data, approaching the experimental value for larger supercells.

\begin{figure}
 \centering
 \includegraphics[width=0.9\columnwidth]{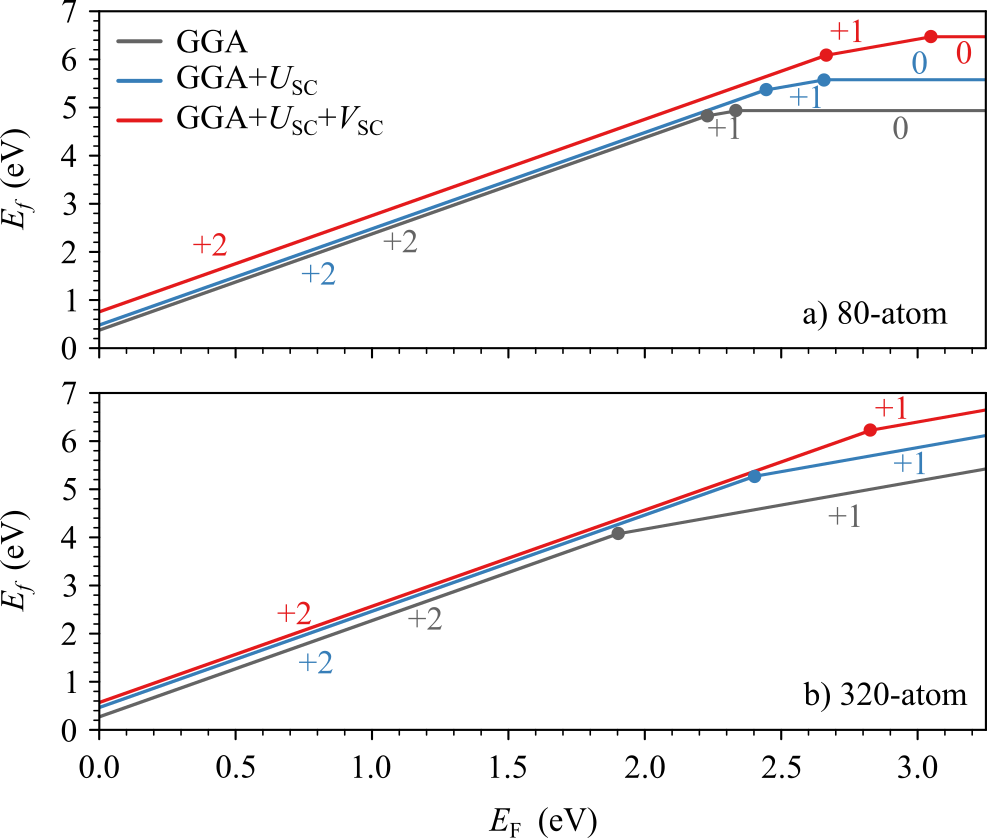}
 \caption{Oxygen vacancy formation energy ($E_f$) in different charge states computed as a function of the Fermi energy ($E_\textrm{F}$, up to the experimental band gap) with respect to the valence band maximum of stoichiometric STO and for $\Delta \mu(O) = 0$ in the a) 80- and b) 320-atom AFD cell. Only the most stable charge state is reported as indicated next to each line. The filled circles denote the transition levels between different charge states computed with equation~\ref{eq:translev}.}
\label{fig:ef_vs_EF_cellsize}
\end{figure}

The formation energies of charged oxygen vacancies also depend on the cell size (see Fig.~\ref{fig:ef_charged}). This dependence is strongest at the GGA level, reflecting again the incorrect electronic structure obtained with this functional. The dependence gets weaker for DFT+$U$, while for DFT+$U$+$V$ it is always very small. The cell-size dependence is reduced when going from V$_\textrm{O}^{\bullet}$ to V$_\textrm{O}^{\bullet \bullet}$, as expected due to the electronic structure of these vacancies for which one or no electron reside in the defect band. Naturally, these results affect also the energetic ordering of the V$_{\textrm{O}}$ as a function of the Fermi energy. As we see from Fig.~\ref{fig:ef_vs_EF_cellsize}a and in line with previous reports~\cite{Mitra2012}, the stability range of the V$_\textrm{O}^{\bullet \bullet}$ in the 80-atom AFD cell is reduced when going from GGA, to GGA+$U$, and GGA+$U$+$V$ with the V$_\textrm{O}^{\bullet \bullet}$/V$_\textrm{O}^{\bullet}$ transition level being pushed toward the CBM, reflecting the electronic-structure changes discussed above. The same effect is observed with increasing cell size (see Fig.~\ref{fig:ef_vs_EF_cellsize}b): for the 320-atom cell not only is the  V$_\textrm{O}^{\bullet \bullet}$/V$_\textrm{O}^{\bullet}$ transition level no longer observed within the experimental STO band gap, but the V$_\textrm{O}^{X}$/V$_\textrm{O}^{\bullet}$ transition level is now very close to the CBM. As a consequence of the increased localization of the V$_\textrm{O}^{\bullet}$ defect state, the stability range of the V$_\textrm{O}^{X}$ is increased when going from GGA, to GGA+$U$, and GGA+$U$+$V$. In agreement with previous results of Choi \textit{et al.}~\cite{Choi2013} on an even larger 1080-atom cell, the shallow nature of the V$_\textrm{O}^{X}$/V$_\textrm{O}^{\bullet}$ transition level thus suggests double ionization of the V$_\textrm{O}$ in a wide temperature range.

Finally, we can observe how the STO phase affects the results. Fig.~\ref{fig:defect_pos_nVO} shows that the cubic phase is always associated with shallower V$_\textrm{O}^{\bullet \bullet}$ defect levels compared to the AFD phase, independently of the functional or cell size and in agreement with a previous report by Choi \textit{et al.}~\cite{Choi2013}. This can be explained with the smaller band gap of cubic STO compared to AFD STO and the associated smaller energetic cost for accommodating the two excess electrons in more delocalized Ti states. Interestingly, however, very similar $E_f$ for the two phases are obtained when a better description of the band gap is provided by the GGA+$U_{\textrm{SC}}$+$V_{\textrm{SC}}$ method (see Fig.~\ref{fig:ef_nVO_cellsize}).

\subsubsection{Atomic relaxations}

\begin{table}
\caption{Displacements of the two Ti atoms adjacent to a V$_\textrm{O}^{\bullet \bullet}$ defect along the Ti$_1$-$\mathrm{V_O}$-Ti$_2$ direction (see Fig.~\ref{fig:relaxation}) in the AFD phase of STO computed for different supercell sizes with different DFT methods. Negative and positive values correspond to outward and inward relaxations, respectively.}
\begin{tabular*}{\columnwidth}{@{\extracolsep{\fill}}cccc}
\hline
\hline
Num. of atoms              & GGA           & GGA+$U_{\textrm{SC}}$         & GGA+$U_{\textrm{SC}}$+$V_{\textrm{SC}}$ \\
\hline
40  &  -0.075 & 0.012  & 0.002  \\
80  &  -0.061 & 0.011   & 0.004  \\
320 &  -0.150 & -0.138 & -0.133 \\
\hline
\hline
\end{tabular*}
\label{tbl:relaxations}
\end {table}

\begin{figure}
 \centering
 \includegraphics[width=0.6\columnwidth]{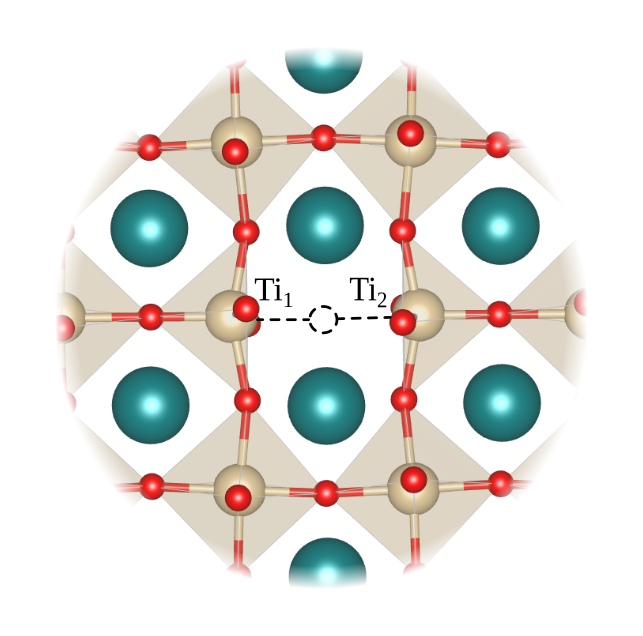}
 \caption{Relaxed structure around the V$_\textrm{O}^{\bullet \bullet}$ defect in a 320-atom cell. The two Ti sites in nearest-neighbor position to the defect are labeled as Ti$_1$ and Ti$_2$.}
\label{fig:relaxation}
\end{figure}

It was suggested that the type of relaxations of the Ti atoms in nearest-neighbor positions to the defect is directly related to the electronic structure of the V$_\textrm{O}^{\bullet \bullet}$ defect~\cite{Mitra2012}. Table~\ref{tbl:relaxations} reports the displacements of these Ti sites along the Ti$_1$-$\mathrm{V_O}$-Ti$_2$ direction (see Fig.~\ref{fig:relaxation}) computed for different cell sizes and with different methods. As was already observed by Mitra \textit{et al.}~\cite{Mitra2012}, at the GGA level the two nearest-neighbor Ti atoms relax away from the defect, independently of the cell size. In this case, no localized states are observed in the gap and consequently the two Ti atoms will gain energy from bonding with the surrounding oxygen atoms. Similarly to DFT results obtained with hybrid functionals~\cite{Mitra2012}, the opposite is observed in the 40- and 80-atom cells at the GGA+$U_{\textrm{SC}}$ and GGA+$U_{\textrm{SC}}$+$V_{\textrm{SC}}$ levels, when shallow but localized defect states are formed in the gap. Indeed, in these cases, the two Ti ions will gain energy by moving towards the two excess electrons trapped in the vacancy site and hence towards each other. Interestingly, for the 320-atom cell (see Fig.~\ref{fig:relaxation}), the two Ti atoms always relax away from each other and by a much larger amount, which agrees with the observation that for larger cells no localized defect states appear in the gap upon V$_\textrm{O}^{\bullet \bullet}$ formation, even when the GGA+$U_{\textrm{SC}}$+$V_{\textrm{SC}}$ method is used. 

\subsubsection{Magnetism}
\begin{figure}
 \centering
 \includegraphics[width=0.9\columnwidth]{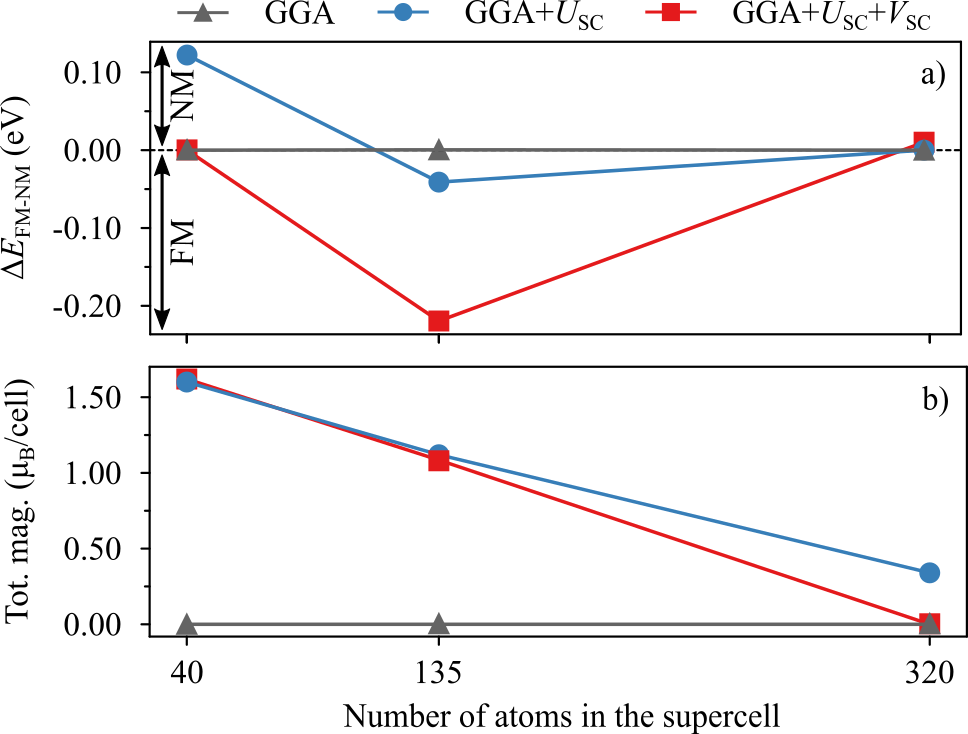}
 \caption{a) Total energy difference ($\Delta E_\textrm{FM-NM}$) between the ferromagnetic (FM) and the non-magnetic (NM) state of a V$_\textrm{O}^{\bullet \bullet}$ in cubic STO cells of different size and computed with different methods (NM is more stable for positive and FM for negative $\Delta E_\textrm{FM-NM}$) and b) total magnetization of the FM state computed with different methods and for supercells of different size.}
\label{fig:magnetization}
\end{figure}
In order to address the effect of spin polarization on the description of the neutral oxygen vacancy, we compare the results for the non-magnetic (NM, singlet) solution for V$_\textrm{O}^{\bullet \bullet}$ reported in the previous sections, with results obtained for a ferromagnetic (FM, triplet) state of the same defect. For comparison with the data reported by Hou and Terakura~\cite{Hou2010}, Fig.~\ref{fig:magnetization}a shows the total energy difference between the FM and NM solutions for a V$_\textrm{O}^{\bullet \bullet}$ in the cubic phase in 40-, 135-, and 320-atom cells, but similar results were obtained for the AFD phase.  GGA is never able to stabilize the FM solution as can be seen also from Fig.~\ref{fig:magnetization}b showing the total magnetization of the cell. Using GGA+$U_{\textrm{SC}}$ and GGA+$U_{\textrm{SC}}$+$V_{\textrm{SC}}$, the FM V$_\textrm{O}^{\bullet \bullet}$ is found to be more stable for the 135-atom cell (by about -0.05 and -0.20 eV, respectively) with a total magnetization of about 1.1 $\mu_{\textrm{B}}$, in good agreement with the results by Hou and Terakura~\cite{Hou2010}. However, both the magnetization and the stability of the FM solution decrease with increasing cell size, the NM and FM solutions having nearly the same energy for the 320-atom cell. This result slightly differs from the one of Hou and Terakura~\cite{Hou2010} who reported the FM solution to be more stable even for this cell size at the DFT+$U$ level of theory. However, in their case, the FM is enforced by fixing the Ti atoms adjacent to the defect at a specific distance from the defect and optimizing only the remaining atomic positions, while in our case no constraints were imposed and all atomic coordinates were allowed to relax.

\begin{figure}
 \centering
 \includegraphics[width=0.9\columnwidth]{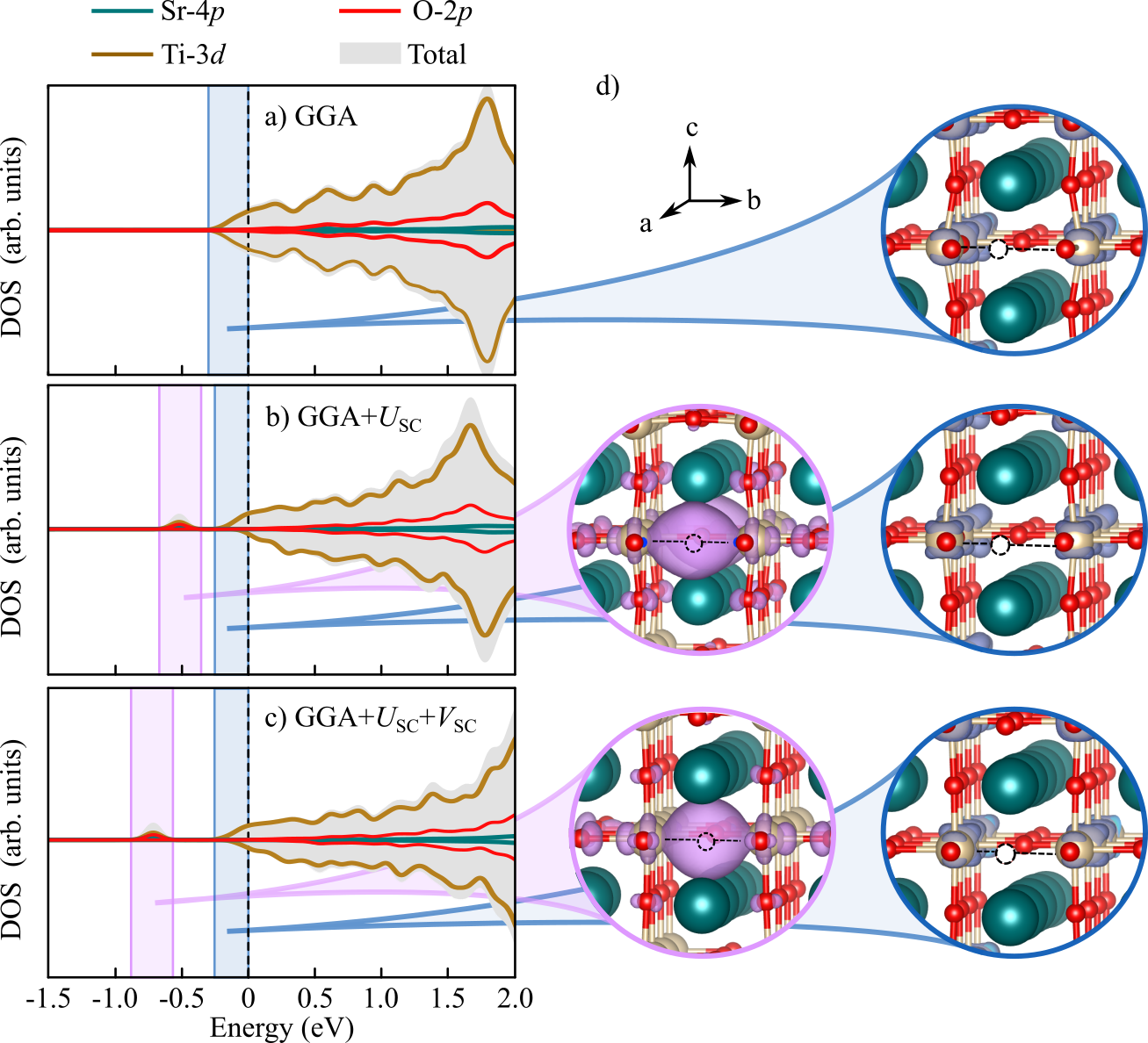}
 \caption{PDOS for a V$_\textrm{O}^{\bullet \bullet}$ in the FM (triplet) state computed with a) GGA, b) GGA+$U_{\textrm{SC}}$, and c) GGA+$U_{\textrm{SC}}$+$V_{\textrm{SC}}$. The vertical dotted line indicates the position of the Fermi level. d) The isosurfaces ($10^{-2} \mathrm{e/\AA}^3$) shown in the circles correspond to the charge density associated with the defect states highlighted with the corresponding color in plots a-c).}
\label{fig:pdos_nVO_fm}
\end{figure}

The destabilization of the FM state for larger supercells can be explained from the electronic properties. Figure~\ref{fig:pdos_nVO_fm} shows the DOS for the FM state in the 135-atom cell. As for the NM case (Fig. \ref{fig:pdos_nVO}), at the GGA level the defect state is fully merged with the CB and the FM solution is not stable. GGA+$U_{\textrm{SC}}$ and GGA+$U_{\textrm{SC}}$+$V_{\textrm{SC}}$ provide instead a different picture in which one electron is localized in an in-gap state with $e_g$ character while the other electron occupies a delocalized $t_{2g}$ state in the CB. As observed for the NM solution, GGA+$U_{\textrm{SC}}$+$V_{\textrm{SC}}$ results in a deeper singly-occupied $e_g$ state. However, also with this approach, this state becomes increasingly shallower with increasing cell size, ultimately accommodating the two electrons in $t_{2g}$ states at the bottom of the CB in the 320-atom cell.

\subsubsection{Self-consistent site-dependent Hubbard parameters}\label{sec:SC-SD}

Defect formation in transition metal oxides can induce local perturbations of the chemical environment of Hubbard sites around the defect, upon which the Hubbard parameters physically depend. For this reason, we recently suggested~\cite{Ricca2019} a self-consistent, site-dependent DFT+$U_{\textrm{SC-SD}}$ approach in which the $U$ values are determined for all inequivalent Hubbard sites. The same procedure can be extended also to DFT+$U$+$V$. This site-dependent approach was found to be promising when a defect is associated with (occupied) deep or well localized states in the band gap, since this localization leads to strong chemical changes for sites around the defect that are properly captured by site-dependent $U$ values.

\begin{figure}
 \centering
 \includegraphics[width=0.9\columnwidth]{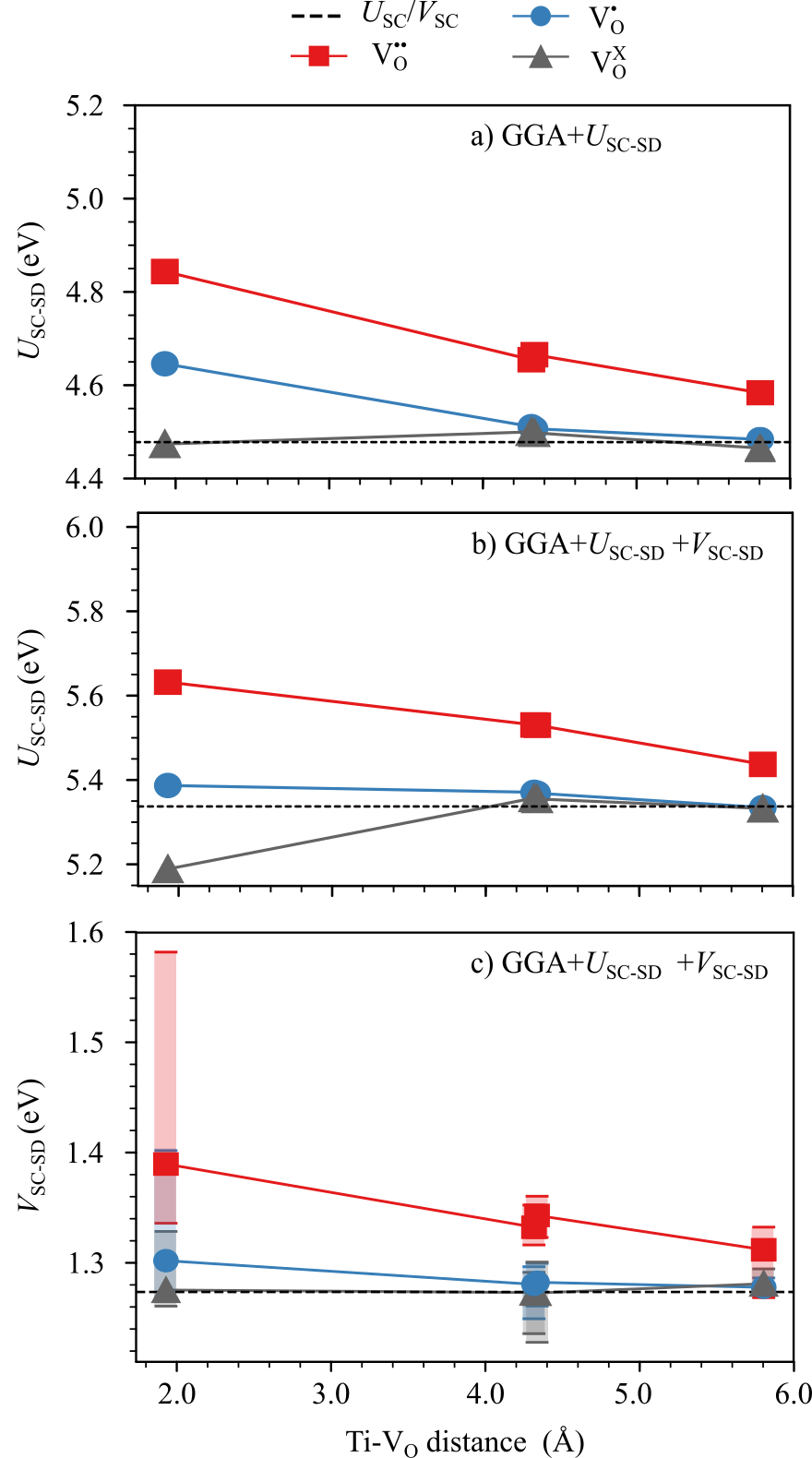}
 \caption{Changes of $U_{\textrm{SC-SD}}$ for Ti sites in the 40-atom AFD cell depending on their distance from V$_\textrm{O}^{\bullet \bullet}$, V$_\textrm{O}^{\bullet}$ and V$_\textrm{O}^{X}$ defects for a) $U$ of GGA+$U$, b) $U$ of GGA+$U$+$V$, and c) $V$ of GGA+$U$+$V$ for the Ti(3$d$)-O(2$p$) interactions averaged over TiO$_6$ octahedra at different distances from the defect. The colored bars indicate the range of $V_{\textrm{SC-SD}}$ at each distance. The dashed horizontal lines indicate the respective SC Hubbard parameters for stoichiometric STO.}
\label{fig:UV_scsd}
\end{figure}

Oxygen vacancies in STO offer the possibility to test the site-dependent approach for defects associated with shallow or band-like states. The $U_{\textrm{SC-SD}}$ values (Fig.~\ref{fig:UV_scsd}a) for a DFT+$U_{\textrm{SC-SD}}$ calculation together with the $U_{\textrm{SC-SD}}$ (Fig.~\ref{fig:UV_scsd}b) and $V_{\textrm{SC-SD}}$ values (Fig.~\ref{fig:UV_scsd}c) obtained within DFT+$U_{\textrm{SC-SD}}$+$V_{\textrm{SC-SD}}$ are shown as a function of the distance of the Hubbard site from a V$_\textrm{O}^{\bullet \bullet}$, V$_\textrm{O}^{\bullet}$, and V$_\textrm{O}^{X}$ defect in a 40-atom AFD cell. For $V_{\textrm{SC-SD}}$, we report the average value together with the minimum and maximum values for the Ti--O pairs in each TiO$_6$ octahedron. SC-SD Hubbard parameters very close to the stoichiometric value are observed for V$_\textrm{O}^{X}$, which is due to the small chemical changes associated with the formation of this defect, consisting of the removal of an O$^{2-}$ anion. Slightly larger deviations on the Ti ions closest to the vacancy (+0.2 eV for $U$ in DFT+$U_{\textrm{SC-SD}}$ and less than +0.1 eV for $U/V$ values in DFT+$U_{\textrm{SC-SD}}$+$V_{\textrm{SC-SD}}$) are generally observed for the V$_\textrm{O}^{\bullet}$, where one O atom and one electron are removed, but the values of the stoichiometric STO system are recovered for larger distances.

Unsurprisingly, the largest changes are obtained when two excess electrons are present after V$_\textrm{O}^{\bullet \bullet}$ formation. The Ti sites in nearest-neighbor positions to the defect show the largest deviations in $U$. The same holds for the $V$ values of Ti(3$d$)-O(2$d$) pairs associated with these two Ti sites showing the largest deviations for the interactions with the O atoms closest to the defect. However, for V$_\textrm{O}^{\bullet \bullet}$, the Hubbard parameters never recover the value of stoichiometric STO even for sites far from the defect.
\begin{figure}
 \centering
 \includegraphics[width=0.9\columnwidth]{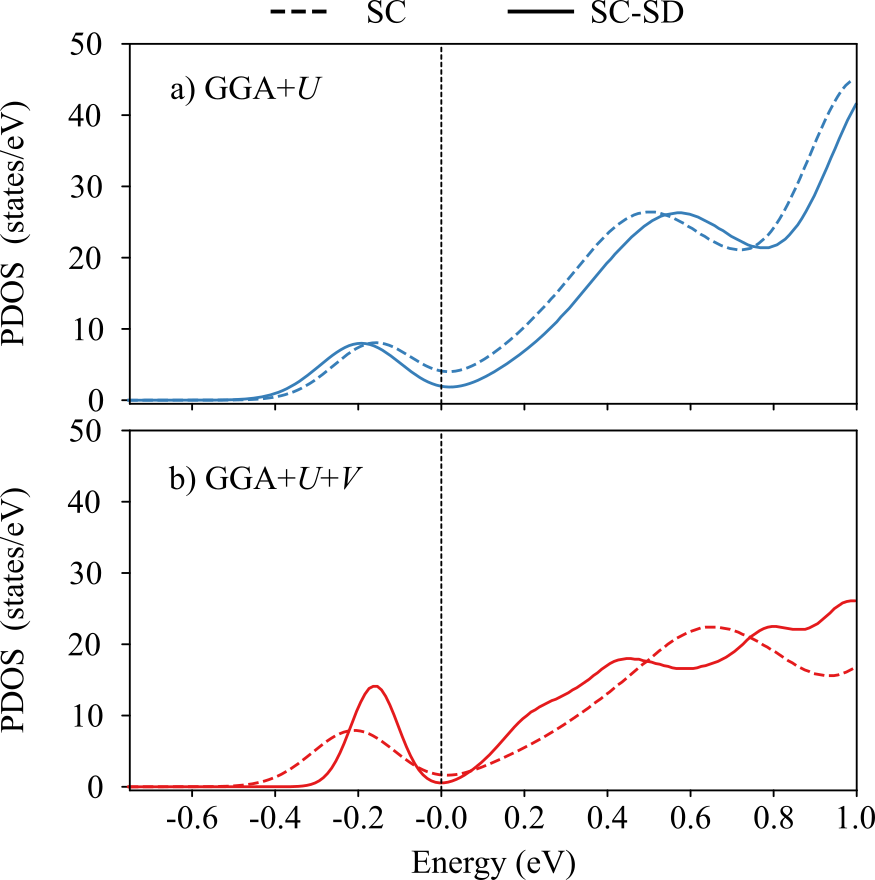}
 \caption{Comparison of the DOS for a V$_\textrm{O}^{\bullet \bullet}$ in the 80-atom cell of the AFD STO phase obtained using SC or SC-SD Hubbard parameters in the case of a) GGA+$U$ and b) GGA+$U$+$V$ calculations. The vertical dotted line indicates the position of the Fermi level.}
\label{fig:pdos_nVO_scsd}
\end{figure}
We believe this behavior to be caused by the peculiar electronic properties of V$_\textrm{O}^{\bullet \bullet}$ in STO, which induce shallow states close to and overlapping with the CB. While the site-dependent Hubbard parameters reduce the dispersion of the defect state (see Fig.~\ref{fig:pdos_nVO_scsd}), it still overlaps with the CB, resulting in the observed long-range dependence of the SC-SD Hubbard parameters.  It is known that increasing the cell size reduces the dispersion of the defect state and its overlap with the CB. However, using larger supercells improved the situation only when going from the 40- to the 80-atom cell (Fig.~\ref{fig:UV_scsd_size}) while for larger cells larger variations reappear, due to the increasingly shallower defect state reported above.

\begin{figure}
 \centering
 \includegraphics[width=0.9\columnwidth]{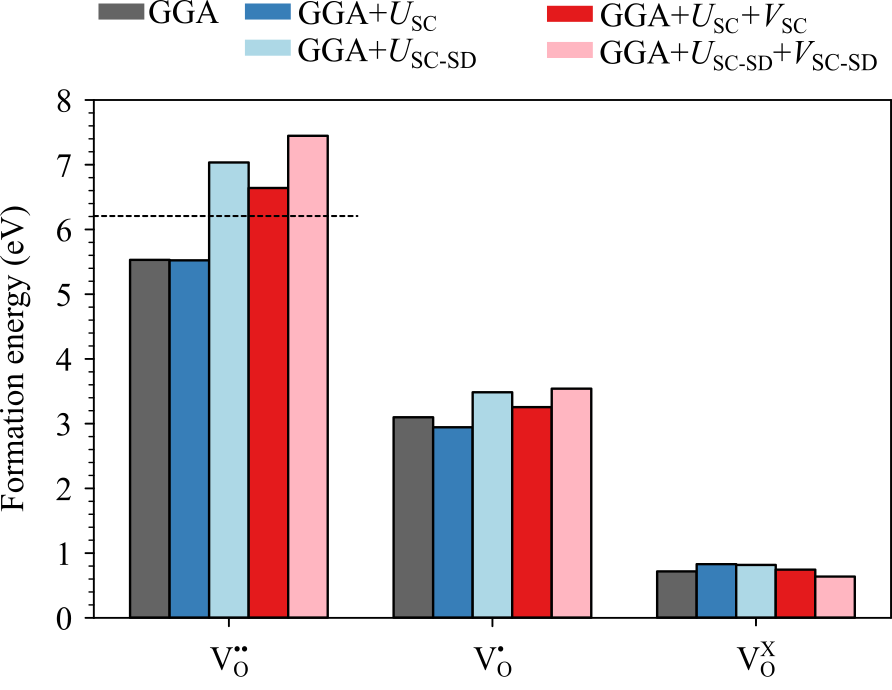}
 \caption{Formation energies computed for V$_\textrm{O}^{\bullet \bullet}$, V$_\textrm{O}^{\bullet}$ and V$_\textrm{O}^{X}$ in a 80-atom AFD STO cell using different methods and for $\Delta \mu(\mathrm{O}) = 0$ and $E_\textrm{F} = 0$. The horizontal dotted line indicates the experimetal value for the neutral defect~\cite{Evarestov2006,Evarestov2012}.}
\label{fig:barchart_ef}
\end{figure}
Figure~\ref{fig:barchart_ef} shows the effect of the SC-SD Hubbard parameters on the computed V$_\textrm{O}$ formation energies. For the V$_\textrm{O}^{X}$, we observe an almost negligible reduction in $E_f$ with respect to the value computed with SC Hubbard parameters of stoichiometric STO, which is a consequence of the small change in the $U/V_{\textrm{SC-SD}}$ parameters compared to the stoichiometric case (Fig. \ref{fig:UV_scsd}). In line with the larger $U/V_{\textrm{SC-SD}}$ changes on the nearest-neighbor sites for the V$_\textrm{O}^{\bullet}$, an increase of about +0.5 and +0.3 eV at the DFT+$U$ or DFT+$U$+$V$ levels of theory respectively is observed. Finally, the long-range site-dependence in the case of the V$_\textrm{O}^{\bullet \bullet}$, results in an unphysical increase of more than 1 eV in the computed $E_f$. A possible explanation could be that the localized atomic orbitals we project on in our DFT+$U$(+$V$) scheme cannot properly capture a shallow F-center defect state. In the present scheme we apply Hubbard corrections only on Ti-3$d$ sites, neglecting the vacancy site where the defect charge mainly localizes. This also explains why this unphysical behavior was not observed in our previous work on V$_\textrm{O}$ in \ce{SrMnO3}~\cite{Ricca2019}, where extra electrons where localized on transition metal sites adjacent to the defect and changes in $U_{\textrm{SC-SD}}$ can properly account for the chemical changes of these sites. These artificial long-range effects could therefore potentially be avoided by alternative, currently not implemented, basis functions for the Hubbard manifold such as maximally localized Wannier functions~\cite{Marzari2012Maximally} that were previously proposed for this purpose~\cite{ORegan2010Projector, Korotin2012Electronic}.

\section{\label{sec:concls}Conclusions}

In this work we investigated the properties of stoichiometric and oxygen-deficient \ce{SrTiO3} (STO) using DFT+$U$+$V$ with Hubbard parameters computed self-consistently via DFPT and compared the results with data obtained within standard DFT, DFT+$U$, and hybrid functional approaches. We find that DFT+$U_{\textrm{SC}}$+$V_{\textrm{SC}}$ yields an improved description of the electronic structure of the stoichiometric AFD phase of STO, with a band gap and $t_{2g}-e_g$ splitting in excellent agreement with experiments. As a consequence, the description of the electronic properties of oxygen vacancies in STO is improved, with the extended Hubbard functional providing formation energies in good agreement with experiments and an overall picture similar to results obtained with hybrid functionals, but at a fraction of the computational cost. Furthermore, the self-consistent determination of the Hubbard parameters not only ensures the internal consistency of our results, but avoids the problem of empirically tuning the fraction of exact exchange in the hybrid functional that strongly affects the predicted defect position and formation energy. We further show that when the defect induces shallow or band-like states, as for a V$_\textrm{O}^{\bullet \bullet}$ in STO, taking into account the site-dependence of Hubbard parameters should be carefully evaluated, as an artificial long-range dependence of these parameters can occur. We believe that alternative functions, such as Wannier functions, for the localized Hubbard manifold could alleviate this issue.

Finally, using a consistent set of calculations, we show how the contradictory theoretical and experimental results for oxygen vacancies in STO can be rationalized in terms of the cell size, phase, magnetic order, atomic relaxations and the exchange-correlation functional.

\section*{\label{sec:acknow}Acknowledgments}

This research was supported by the Swiss National Science Foundation (SNSF), through Grant No. 200021-179138, and its National Centre of Competence in Research (NCCR) MARVEL. Computational resources were provided by the University of Bern (on the HPC cluster UBELIX, http://www.id.unibe.ch/hpc), by the Swiss National Supercomputing Center (CSCS) under projects ID mr26 and s836 and SuperMUC at GCS@LRZ, Germany, for which we acknowledge PRACE for awarding us access.

\bibliography{references}


\clearpage
\clearpage 
\setcounter{page}{1}
\renewcommand{\thetable}{S\arabic{table}}  
\setcounter{table}{0}
\renewcommand{\thefigure}{S\arabic{figure}}
\setcounter{figure}{0}
\renewcommand{\thesection}{S\arabic{section}}
\setcounter{section}{0}
\renewcommand{\theequation}{S\arabic{equation}}
\setcounter{equation}{0}
\onecolumngrid

\begin{center}
\textbf{Supplementary information for\\\vspace{0.5 cm}
\large Self-consistent DFT+$U$+$V$ study of oxygen vacancies in SrTiO$_3$\\\vspace{0.3 cm}}
Chiara Ricca,$^{1}$ Iurii Timrov,$^{2}$ Matteo Cococcioni,$^{2, 3}$ Nicola Marzari,$^{2}$ and Ulrich Aschauer$^{1}$

\small

$^1$\textit{Department of Chemistry and Biochemistry and National Centre for Computational Design and Discovery of Novel Materials (MARVEL), University of Bern, Freiestrasse 3, CH-3012 Bern, Switzerland}

$^2$\textit{Theory and Simulation of Materials (THEOS) and National Centre for Computational Design and Discovery of Novel Materials (MARVEL), Ecole Polytechnique F\'ed\'erale de Lausanne, CH-1015 Lausanne, Switzerland}

$^3$\textit{Department of Physics, University of Pavia, Via A. Bassi 6, 27100 Pavia, Italy}

(Dated: \today)
\end{center}

\section{Stoichiometric STO}

\subsection{\label{secSI:structure_STO}Structure}

\begin{table}[b]
\caption{Comparison of the calculated and experimental structural properties (lattice parameter \textit{a} in \angstrom, 
 \textit{c}/\textit{a} ratio, and octahedral rotation angle around the $c$-axis $\theta $ in degrees), and band gap ($E_\textrm{g}$, in eV) of STO in the AFD and cubic phases. When applicable, the corresponding self-consistent $U$ and average $V$ values are reported.}
\small
\begin{tabular}{llcccccc}
\hline
\hline
Supercell                     & Method      & $U_\mathrm{SC}$     & $V_\mathrm{SC}$     & $a$     & $c$/$a$   & $\theta$ & $E_\mathrm{g}$    \\
\hline
\multirow{3}{*}{40-atom AFD}  & GGA         & - & - & 3.860 & 1.006 & 5.69  & 2.00     \\
                              & GGA+$U_\mathrm{SC}$     & 4.48  & -     & 3.857  & 1.012 & 7.81  & 2.62  \\
                              & GGA+$U_\mathrm{SC}$+$V_\mathrm{SC}$ & 5.34  & 1.27  & 3.841  & 1.009 & 6.47  & 3.11  \\
\hline
\multirow{3}{*}{80-atom AFD}  & GGA         & -     & -     & 3.855  & 1.008 & 6.36  & 2.03  \\
                              & GGA+$U_\mathrm{SC}$     & 4.48  & -     & 3.857  & 1.012 & 7.83  & 2.61  \\
                              & GGA+$U_\mathrm{SC}$+$V_\mathrm{SC}$ & 5.30  & 1.26  & 3.841  & 1.009 & 6.52  & 3.09  \\
\hline
\multirow{3}{*}{320-atom AFD} & GGA         & -     & -     & 3.858  & 1.007 & 6.42  & 2.02  \\
                              & GGA+$U_\mathrm{SC}$     & 4.48  & -     & 3.857  & 1.013 & 7.93  & 2.60   \\
                              & GGA+$U_\mathrm{SC}$+$V_\mathrm{SC}$ & 5.37  & 1.31  & 3.840  & 1.010 & 6.62  & 3.11  \\
\hline
                              & Exp.        & -     & -     & 3.898$^a$   & 1.006$^a$     & 2.1$^b$   & -  \\
\hline
\hline
\multirow{3}{*}{40-atom Cubic}    & GGA         & -     & -     & 3.870  & -     & -     & 1.88  \\
                              & GGA+$U_\mathrm{SC}$     & 4.45  & -     & 3.877  & -     & -     & 2.43  \\
                              & GGA+$U_\mathrm{SC}$+$V_\mathrm{SC}$ & 5.35  & 1.31  & 3.855  & -     & -     & 2.97  \\
\hline
\multirow{3}{*}{80-atom Cubic}    & GGA         & -     & -     & 3.870  & -     & -     & 1.88 \\
                              & GGA+$U_\mathrm{SC}$     & 4.44  & -     & 3.878  & -     & -     & 2.44  \\
                              & GGA+$U_\mathrm{SC}$+$V_\mathrm{SC}$ & 5.27  & 1.25  & 3.858  & -     & -     & 2.94  \\
\hline
\multirow{3}{*}{135-atom Cubic}   & GGA         & -     & -     & 3.870  & -     & -     & 1.92  \\
                              & GGA+$U_\mathrm{SC}$     & 4.45  & -     & 3.878  & -     & -     & 2.47 \\
                              & GGA+$U_\mathrm{SC}$+$V_\mathrm{SC}$ & 5.32  & 1.28  & 3.856  & -     & -     & 3.00 \\
\hline
\multirow{3}{*}{320-atom Cubic}   & GGA         & -     & -     & 3.870  & -     & -     & 1.86  \\
                              & GGA+$U_\mathrm{SC}$     & 4.45  & -     & 3.879  & -     & -     & 2.41  \\
                              & GGA+$U_\mathrm{SC}$+$V_\mathrm{SC}$ & 5.34  & 1.30   & 3.856  & -     & -     & 2.94 \\
\hline
                              & Exp.        & -     & -     & 3.900$^a$   & -     & -     & 3.25$^c$  \\
\hline
\hline
\end{tabular}
\begin{flushleft}
\small
$^a$ Ref.~\citeSI{Cao2000_SI} data at 65 K for the AFD phase of STO.\\
$^b$ Ref.~\citeSI{Unoki1967_SI} data at 4.2 K.\\
$^c$ Ref.~\citeSI{vanBenthem2001_SI} at room temperature. \\
\end{flushleft}
\label{tbl:SI_sto_structure}
\end {table}
\setlength{\parskip}{1em}
Table~\ref{tbl:SI_sto_structure} shows the structural parameters computed for the AFD and the cubic STO phases with different DFT methods, together with the corresponding self-consistent Hubbard parameters calculated by DFPT. Very similar $U_\mathrm{SC}$ and $V_\mathrm{SC}$ values are obtained for the different cell sizes in both phases, reflecting a proper convergence of these quantities with respect to the \textbf{q}-mesh applied in each case. In particular, the Hubbard parameters are consistently  smaller for the AFD phase, probably due to the different crystal environment of the Ti due to octahedral rotations. Consistent structural results are obtained for the different cell sizes: in the cubic phase, GGA+$U_\mathrm{SC}$ provides the best agreement with experiments, while GGA is associated with the smallest errors for the AFD phase. In all cases GGA+$U_\mathrm{SC}$+$V_\mathrm{SC}$ results in the smallest $a$, because the inter-site interactions encourage the occupations of hybridized states, shortening the bonds. However, for the AFD phase GGA+$U_\mathrm{SC}$+$V_\mathrm{SC}$ reduces the error on the $c/a$ ratio and rotation angle with respect to GGA+$U_\mathrm{SC}$. Band gaps are predicted similarly for different cell sizes, with values in the cubic phase being consistently smaller than in the AFD phase. This is a consequence of the octahedral rotations that decrease the band width. While GGA and GGA+$U_\mathrm{SC}$ provide underestimated band gaps, GGA+$U_\mathrm{SC}$+$V_\mathrm{SC}$ is in excellent agreement with experiments.

It is important to note that the underestimation of the lattice parameter $a$ at the GGA+$U_\mathrm{SC}$+$V_\mathrm{SC}$ level is a result of the underlying GGA functional. Table~\ref{tbl:SI_structure_pbe} shows a comparison of the Hubbard parameters and resulting $a$ values for the 5-atom unit cell of cubic STO obtained with the PBEsol (as in Table~\ref{tbl:SI_sto_structure}) and the PBE GGA functionals. As expected PBEsol results in a smaller lattice parameter compared to PBE, which, instead, overestimates $a$ compared to experiment.

Consequently, since GGA+$U_\mathrm{SC}$ expands $a$ with respect to the GGA results, the previously underestimated $a$ for the PBEsol functional approaches experiment with the GGA+$U_\mathrm{SC}$ correction, while the already overestimated $a$ for the PBE functional deviates further from experiment. Interestingly, self-consistent Hubbard $U_\mathrm{SC}$ and $V_\mathrm{SC}$ parameters are quite different with these two functionals: The more ``compressed'' PBEsol structure seems to favor covalent bonding, resulting in a $V_\mathrm{SC}$ value about 0.37~eV larger than in the PBE structure. Hence, while in the PBE+$U_\mathrm{SC}$+$V_\mathrm{SC}$ case, the tendency of $U$ to expand and of $V$ to contract the lattice are somewhat compensated, resulting in a $a$ value close to the GGA data, for PBEsol the larger $V_\mathrm{SC}$ is responsible for the reduction in lattice parameter compared to plain PBEsol.
\begin{table}[h]
\caption{Comparison of the calculated and experimental lattice parameter (\textit{a} in \angstrom), and band gap ($E_\textrm{g}$, in eV) of cubic STO. Results were obtained for the 5-atom unit cell
with the PBEsol and the PBE functionals and different DFT methods. When necessary, the corresponding self-consistent $U$ and average $V$ values are also reported. We note here that discrepancies between PBEsol results in Tables~\ref{tbl:SI_sto_structure}
. and~\ref{tbl:SI_structure_pbe} are due to numerical differences between the calculations for the 5- and 40-atom cells.}
\small
\begin{tabular}{llcccc}
\hline
\hline
Functional                     & Method      & $U_\mathrm{SC}$     & $V_\mathrm{SC}$     & $a$    &  $E_\mathrm{g}$ \\
\hline
PBEsol  	  & GGA         & - & - & 3.864 &    1.89     \\
                              & GGA+$U_\mathrm{SC}$     & 4.45  & -     & 3.877    & 2.43  \\
                              & GGA+$U_\mathrm{SC}$+$V_\mathrm{SC}$ & 5.35  & 1.31  & 3.855    & 2.97 \\
\hline
\multirow{3}{*}{PBE}  	  	  & GGA         & - & - & 3.938 &   1.83     \\
                              & GGA+$U_\mathrm{SC}$     & 4.74  & -     & 3.958    & 2.36  \\
                              & GGA+$U_\mathrm{SC}$+$V_\mathrm{SC}$ & 5.27  & 0.94  & 3.943    & 2.74 \\
\hline
                              & Exp.        & -     & -     & 3.900$^a$     & 3.25$^b$  \\
\hline
\hline
\end{tabular}
\begin{flushleft}
\small
$^a$ Ref.~\citeSI{Cao2000_SI} data at 65 K for the AFD phase of STO.\\
$^b$ Ref.~\citeSI{vanBenthem2001_SI}. \\
\end{flushleft}
\label{tbl:SI_structure_pbe}
\end {table}
\setlength{\parskip}{0em}
\subsection{Effect of the fraction of exact exchange on the electronic properties}
\begin{figure}[h]
 \centering
 \includegraphics[width=0.35\textwidth]{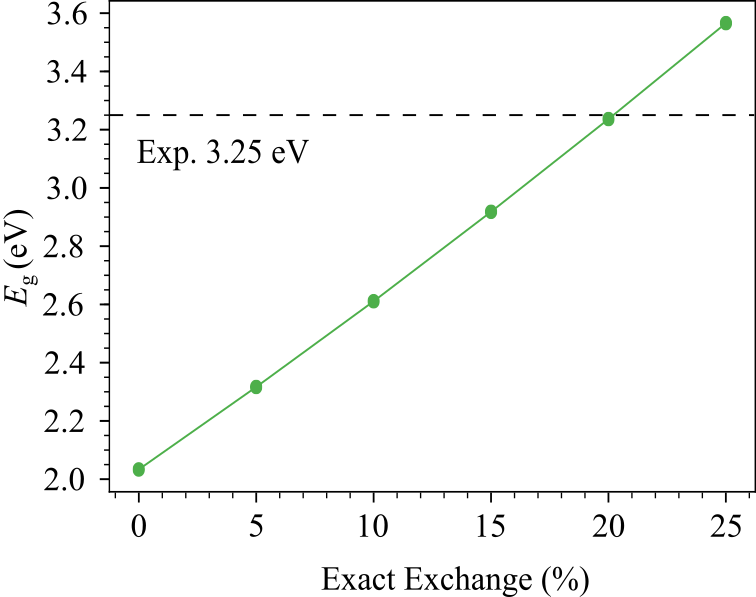}
 \caption{Band gap for the AFD phase of STO computed in a 40-atom supercell using the hybrid HSE functional while varying the fraction of exact exact exchange.}
\label{fig:STO_bandgap_hybrid}
\end{figure}

The percentage of exact exchange in a hybrid functional is a material and property dependent parameter, as can be seen for the STO band gap in Fig.~\ref{fig:STO_bandgap_hybrid}. The band gap linearly depends on the percentage of exact exchange in the HSE functional. HSE06 (25\% of exact exchange), often used for STO~\citeSI{ElMellouhi2013structural_SI, Choi2013_SI, Mitra2012_SI}, overestimates the band gap by about 0.32~eV with respect to experiment, the best agreement being obtained with about 20\% exact exchange.

\clearpage
\section{Oxygen-deficient STO}
\subsection{Defect Position}
\setlength{\parskip}{1em}
The position of the defect state associated with a V$_\mathrm{O}^{\bullet \bullet}$ is one of the most debated properties in oxygen-deficient STO. Table~\ref{tbl:defectlevel} compares our results (obtained for different STO phases, supercell sizes, magnetic states and with different DFT methods) with results of previous theoretical and experimental works.

 Theoretical studies based on LDA and GGA functionals~\citeSI{Luo2004_SI, DJERMOUNI2010904_SI, Hamid2009_SI, Shanthi1998_SI, Astala200181_SI, Tanaka2003_SI, Mitra2012_SI} always find the two extra electrons occupying the conduction band (CB) of STO, with the exception of Ref.~\citeSI{Astala2001_SI} where a very shallow defect state was reported in the ferromagnetic (FM, triplet) configuration.

DFT+$U$ results instead show more variation. For large cells (from 160 to 1080 atoms)~\citeSI{Hou2010_SI, Choi2013_SI}, especially in the cubic STO phase and for the non magnetic (singlet) case, the two electrons are also found in extended $t_{2g}$ states at the bottom of the CB, similar to results with standard LDA/GGA functionals. The majority of the data reported for smaller cells in the non magnetic (singlet) state report instead a doubly occupied shallow defect state about 0.1~eV below the CB~\citeSI{Hou2010_SI, Tanaka2003_SI, Cuong2007_SI}, which is in good qualitative agreement with our results of 0.09 to 0.60 eV (depending on the cell size and STO phase). Instead, a deeper singly occupied defect state combined with a second electron localized in the CB is found by DFT+$U$ when considering the ferromagnetic (triplet) solution: this state was found to lie 0.4 to 0.7~eV below the CB~\citeSI{Choi2013_SI, Tanaka2003_SI, Hou2010_SI}, which is in agreement with our results of 0.2 and 0.8~eV (depending on the cell size and STO phase). A deeper lying doubly occupied state is reported by DFT+$U$ only when very large $U$ values of about 8 eV are applied (0.7~eV below the CB)~\citeSI{Mitra2012_SI} or when different polaron configurations~\citeSI{Hao2015_SI} are explicitly taken into account (0.8-1.1 ~eV below the CB). 

Hybrid functionals generally lead to deep in-gap states occupied by two electrons and lying 0.7 to 1.1~eV below the CB in the 40-atom cell and becoming slightly shallower (up to 0.4~eV from the CB) with increasing cell size~\citeSI{Evarestov2006_SI, Alexandrov2009_SI, Carrasco2005_SI, Zhukovskii2009_SI, ZHUKOVSKII20091359_SI, ElMellouhi2013structural_SI, Mitra2012_SI}. Very deep defect states at about 1.2~eV were instead predicted by dynamical mean field theory (DMFT) for vacancies at the STO surface in a LaAlO$_3$/SrTiO$_3$ heterostructure~\citeSI{Lecherman2014_SI}. 

Defect-state positions predicted in this work by DFT+$U$+$V$ are generally deeper compared to DFT+$U$ results, but not as deep as predicted by hybrid functionals for similar cell sizes: the doubly occupied shallow defect states is found 0.1 to 0.6~eV below the CB in small STO supercells or in the CB for the large 320-atom cell. 

Unfortunately, the comparison of these results with experiments is not straightforward, since fairly different results were obtained with different techniques and for different types of samples with different defect concentrations. Ionization energies for STO single crystals derived from Hall or electrical conductivity measurements~\citeSI{Tufte1967_SI,Lee1971_SI, Moos1997_SI, Raevski_1998_SI} lie between 0.003 and 0.4~eV, suggesting that the formation of V$_\mathrm{O}^{\bullet \bullet}$ defects in STO crystals is associated with fairly shallow defect states, in line with the DFT+$U$+$V$ results. Similar results were obtained from the analysis of the photoluminescence \citeSI{Kan2005_SI} and of UV-vis (ultra violet-visible)\citeSI{Ravichandran2011_SI} spectra of STO crystals and of the ultraviolet photoemission spectra (UPS)~\citeSI{Henrich1978_SI} of STO surfaces, reporting the defect state at 0.4~eV and 0.1~eV from the CB, respectively. In other cases, much deeper defect states at about 1~eV  were observed instead in the UPS spectra~\citeSI{Henrich1978_SI, Courths1980_SI}. While these results were often used to validate hybrid functional results, they are however obtained for STO surfaces and in some cases after ion (\ce{Ar+}) irradiation to induce defects.
\begin{table}
\caption{Position of the defect state  associated to a V$_\textrm{O}^{\bullet \bullet}$ computed with respect to the CB (dara reported in eV) as obtained with different DFT methods in this work and in previously published theoretical and experimental studies. For each entry, we report the STO phase (cubic or AFD) and for theoretical results, we also indicate  the DFT functional, the basis set (PW for plane waves and LCAO for linear combination of atomic orbitals), the number of atoms in the supercell and the considered magnetic state (NM for the non magnetic singlet and FM for the ferromagnetic triplet solution).}
\footnotesize
\resizebox{\linewidth}{!}{%
\begin{tabular*}{\columnwidth}{@{\extracolsep{\fill}}llllll}
\hline
\hline
Method                                        & Basis Set             & STO phase             & N atoms              & Magnetism           & Defect Position \\
\hline
GGA {[}This work{]}                           & PW					  & Cubic/AFD                 & 40-320               & NM                  & CB              \\
GGA+$U$  {[}This work{]}      				  						  &                       & AFD					  & 40                   & NM				   & 0.60             \\
                                              &                       & AFD                   & 80                   & NM                    & 0.23            \\
                                              &                       & AFD                   & 320                  & NM                    & CB              \\
				      						  &                       & AFD                   & 40                   & FM					 & 0.82            \\
                                              &                       & AFD                   & 80                   & FM                    & 0.54            \\
      				  						  &                       & Cubic  				  & 40                   & NM 					& 0.60             \\
                                              &                       & Cubic                     & 80                   & NM                    & 0.09            \\
                                              &                       & Cubic                     & 320                  & NM                    & CB              \\
        									  &                       & Cubic                     & 40                   & FM 					& 1.25            \\
                                              &                       & Cubic                     & 80                   & FM                   & 0.52            \\
GGA+$U$+$V$ {[}This work{]}     				  &                       & AFD                   & 40                   &    NM                 & 0.65            \\
                                              &                       & AFD                      & 80                   & NM                    & 0.30             \\
                                              &                       & AFD                      & 320                  &  NM                   & CB              \\
      									      &                       & AFD                      & 40                   &  FM                   & 1.03            \\
                                              &                       & AFD                     & 80                   &   FM                & 0.29            \\
    										  &                       & Cubic                     & 40                   &   NM                  & 0.64            \\
                                              &                       & Cubic                      & 80                   &  NM                   & 0.17            \\
                                              &                       & Cubic                     & 320                  & NM                    & CB              \\
     										  &                       & Cubic                     & 40                   &   FM                  & 0.99            \\
                                              &                       & Cubic                    & 80                   & FM                    & 1.78            \\
\hline
LDA~\citeSI{Luo2004_SI}                          & PW                    & Cubic                     & 40-80                & NM                  & CB              \\

LDA~\citeSI{DJERMOUNI2010904_SI}                   & PW                    & Cubic                     & 20,40,60,80          & NM                  & CB              \\
LSDA~\citeSI{Hamid2009_SI}                         & PW                    & Cubic                    & 40                   & NM                  & CB              \\
LSDA~\citeSI{Shanthi1998_SI}                     & LMTO-ASA$^a$              & Cubic                     & 40                   & NM                  & CB              \\
LDA~\citeSI{Mitra2012_SI}                        & PW                    & Cubic                     & 80                   & NM                  & CB              \\
LDA~\citeSI{Astala200181_SI}         			  & PW                    & Cubic                     & 40                   & NM                  & CB \\
LDA~\citeSI{Tanaka2003_SI}                       & PW                    & Cubic                     & 135                  & NM                  & 0.08            \\
LSDA~\citeSI{Astala2001_SI}         			  & PW                    & Cubic                     & 40                   & FM                  & $~$0.1            \\
LDA+$U$~\citeSI{Cuong2007_SI}$^b$                    & PW                    & Cubic                     & 320                  & NM                  & 0.11             \\
GGA+$U$~\citeSI{Hou2010_SI}$^b$                    & PW  & Cubic    & 160 & NM                  & CB              \\
                                              &                       &                       & 160                     & NM                  & 0.1             \\
                                              &                       &                       &  160                    & FM                  & 0.5             \\
GGA+$U$ ~\citeSI{Choi2013_SI}$^b$                   & PW   & AFD                   & 135                  & FM                  & 0.5             \\
                                              &                       & Cubic                     & 135                  & FM                  & 0.5             \\
                                              &                       & Cubic                     & 320                  & FM                  & 0.4             \\
                                                                                            &                       &                       & 625-1080             & NM                  & CB              \\
GGA+$U$~\citeSI{Mitra2012_SI}$^c$                               & PW                    & Cubic                     & 80                   & NM                  & 0.7             \\
GGA+$U$~\citeSI{Hao2015_SI}$^d$                 & PW                    & Cubic/AFD                 & 625                  & NM?                 & 0.8-1.1         \\
B3PW~\citeSI{Evarestov2006_SI} & LCAO & Cubic    & 80                   & NM                  & 0.69            \\
                                              &                       &                       & 135                  & NM                  & 0.72            \\
                                              &                       &                       & 160                  & NM                  & 0.57            \\
                                              &                       &                       & 270                  & NM                  & 0.49            \\
                                              &                       &                       & 320                  & NM                  & 0.49            \\
B3LYP~\citeSI{Ricci2003_SI}                                   & LCAO                  & Cubic                     & 80                   & NM                  & 0.8             \\
B3PW~\citeSI{Alexandrov2009_SI}                             & LCAO                  & Cubic                     & 80                   & NM                  & 0.79            \\
B3PW~\citeSI{Carrasco2005_SI}              & LCAO & Cubic    & 160                  & NM                  & 0.77            \\
                                              &                       &                       & 40                   & NM                  & 1.1             \\
                                              &                       &                       & 80                   & NM                  & 0.75            \\
B3PW~\citeSI{Zhukovskii2009_SI,ZHUKOVSKII20091359_SI}             & LCAO & Cubic   & 135                  & NM                  & 0.69            \\

HSE06~\citeSI{Mitra2012_SI}                      & PW                    & Cubic                     & 80                   & NM                  & 0.7             \\
HSE06~\citeSI{ElMellouhi2013structural_SI}            & LCAO & Cubic   & 40                   & NM                  & 0.44            \\
                                              &                       &                       & 90                   & NM                  & 0.42            \\
DFT+DMFT~\citeSI{Lecherman2014_SI}$^{e}$      &                       & Cubic              & -                & PM        & 1.2            \\
\hline
Hall conduction measurements~\citeSI{Tufte1967_SI,Lee1971_SI}$^{f}$                            & -                     & Cubic                     & -                    & -                   & 0.07-0.16       \\
Electrical and Hall conductivity~\citeSI{Moos1997_SI}$^{g}$              & -                     & Cubic                     & -                    & -                   & 0.003, 0.3               \\
Electrical conductivity~\citeSI{Raevski_1998_SI}               & -                     & Cubic                     & -                    & -                   & 0.3-0.4               \\
UV-Vis~\citeSI{Ravichandran2011_SI}$^{h}$, photoluminescence~\citeSI{Kan2005_SI}$^{i}$              & -                     & Cubic                     & -                    & -                   & 0.4             \\
UPS~\citeSI{Henrich1978_SI}$^{l}$                                      & -                     & Cubic                     & -                    & -                   & 0.1 \\
UPS~\citeSI{Henrich1978_SI}$^{m}$                                      & -                     & Cubic                     & -                    & -                   & $~$1.0 \\
UPS~\citeSI{Courths1980_SI}$^{n}$                  & -                     & Cubic                     & -                    & -                   & 1.2             \\
\hline
\hline
\end{tabular*}}
\label{tbl:defectlevel}
\end {table}

\clearpage

\begin{flushleft}
\footnotesize
$^a$ Linearized muffin-tin orbital method within the atomic-sphere approximation.\\
$^b$ $U=$ 5.0 eV and $J=$ 0.64 eV.\\
$^c$ $U=$ 8.0 eV.\\
$^d$ $U=$ 4.96 eV and $J=$ 0.51 eV; different polaron configurations.\\
$^{e}$ V$_\textrm{O}$ at the SrTiO$_3$ surface in a LaAlO$_3$/SrTiO$_3$ heterostructure.\\
$^{f}$ Ionization energy in single crystals for increasing defect concentration.\\
$^{g}$ Ionization energies for V$_\textrm{O}^{\bullet \bullet} \rightarrow$ V$_\textrm{O}^{\bullet}$ and V$_\textrm{O}^{\bullet} \rightarrow$ V$_\textrm{O}^{\textrm{X}}$, respectively.\\
$^{h}$ 15\% La-doped STO.\\
$^{i}$ Photoluminescence spectra of Ar$^+$-irradiated STO crystals.\\
$^{l}$ Ultraviolet photoemission spectra (UPS) of the STO-(100) surface.\\
$^{m}$ UPS of the STO-(100) surface after ion (Ar bombardment) to induce defect formation.\\
$^{n}$ UPS of STO-(100) after Ar sputtering.\\
\end{flushleft}

\setlength{\parskip}{0em}
\subsection{Electronic Properties of Oxygen Defects in STO}
\subsubsection{Effect of the fraction of exact exchange in HSE on the defect properties}

\begin{figure}[h]
 \centering
 \includegraphics[width=0.6\textwidth]{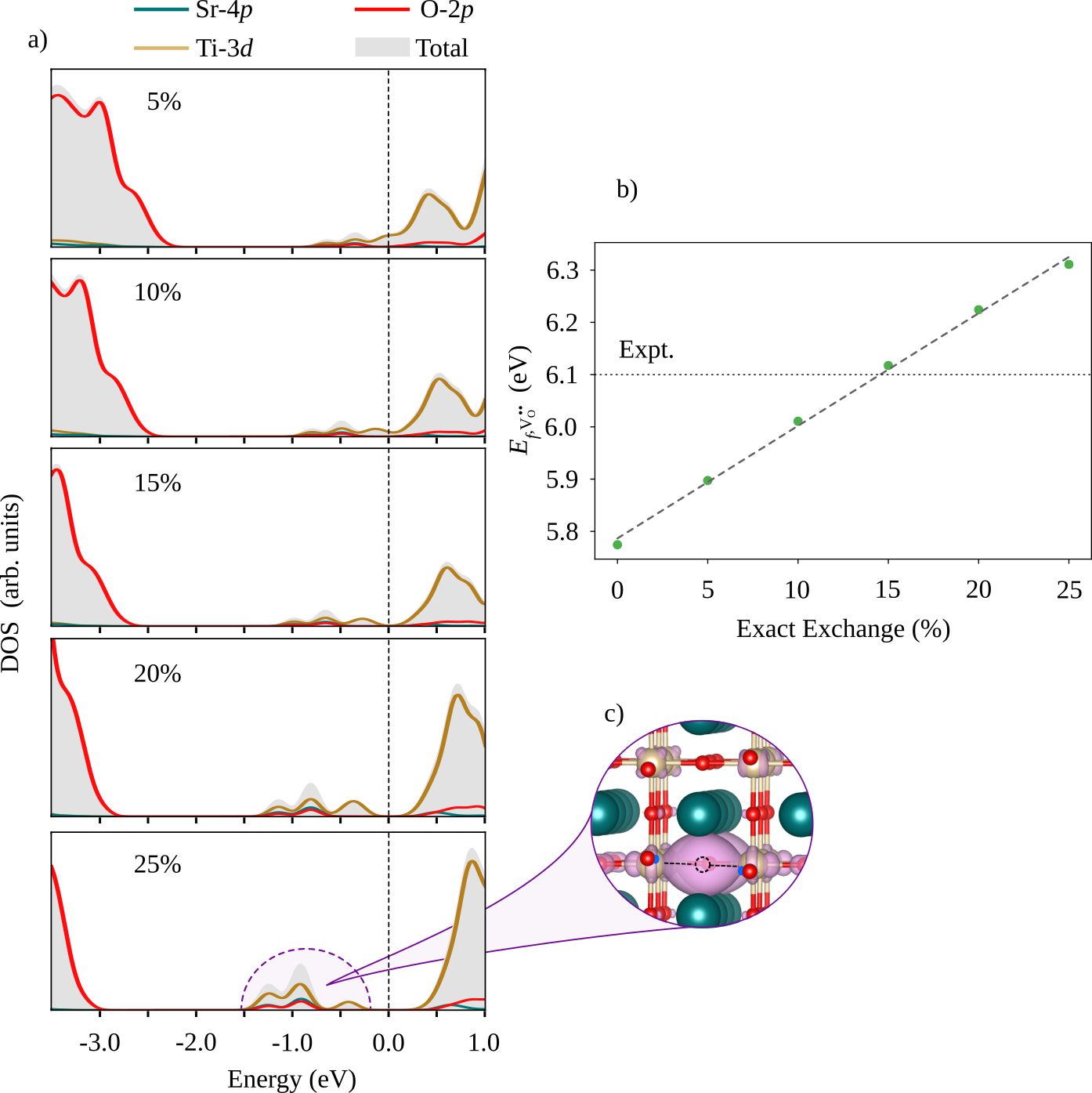}
 \caption{a) PDOS for a V$_\textrm{O}^{\bullet \bullet}$ in a 40-atom AFD cell and b) the computed defect formation energy as a function of the fraction of exact exchange included in the HSE hybrid functional. The vertical dotted line in a) indicates the position of the Fermi level. The isosurface (10$^{-2} e$ \angstrom$^{-3}$) reported in c) corresponds to the density associated with the circled defect states. The vacancy position is indicated by the dashed circle.}
\label{fig:pdos_hse}
\end{figure}

The defect-state position (see Fig.~\ref{fig:pdos_hse}a) and consequently the computed defect formation energy (see Fig.~\ref{fig:pdos_hse}b) depends on the  percentage of exact exchange included into the HSE functional. As expected from the increase in band gap with increasing fraction of exact exchange shown in Fig.~\ref{fig:STO_bandgap_hybrid}, the defect state becomes increasingly deeper with increasing percentage of exact exchange. Consequently, the defect formation energy is also found to increase linearly with the fraction of exchange, due to the larger cost associated with the localization of the two extra electrons in a defect level lying at increasingly  lower energy.

\subsubsection{Electronic properties of charged defects}\label{sec:elecprop_chargedVO}

\begin{figure}[h]
 \centering
 \includegraphics[width=0.5\textwidth]{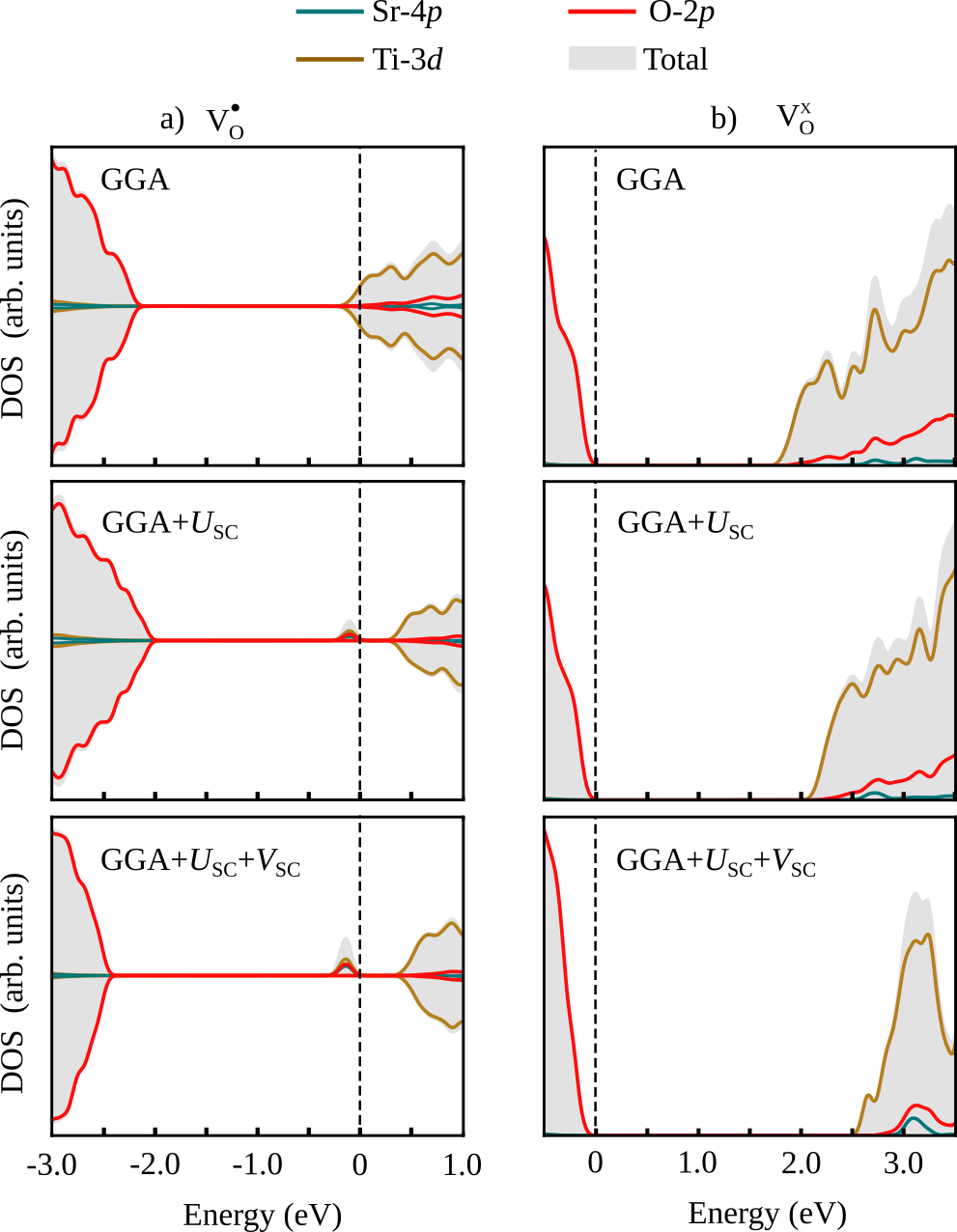}
 \caption{PDOS for a) a V$_\textrm{O}^{\bullet}$ and b) a V$_\textrm{O}^{X}$
 in the 80-atom AFD cell computed with different methods. The vertical dotted line indicates the position of the Fermi level.}
\label{fig:pdos_charged}
\end{figure}

\begin{figure}[h]
 \centering
 \includegraphics[width=0.41\textwidth]{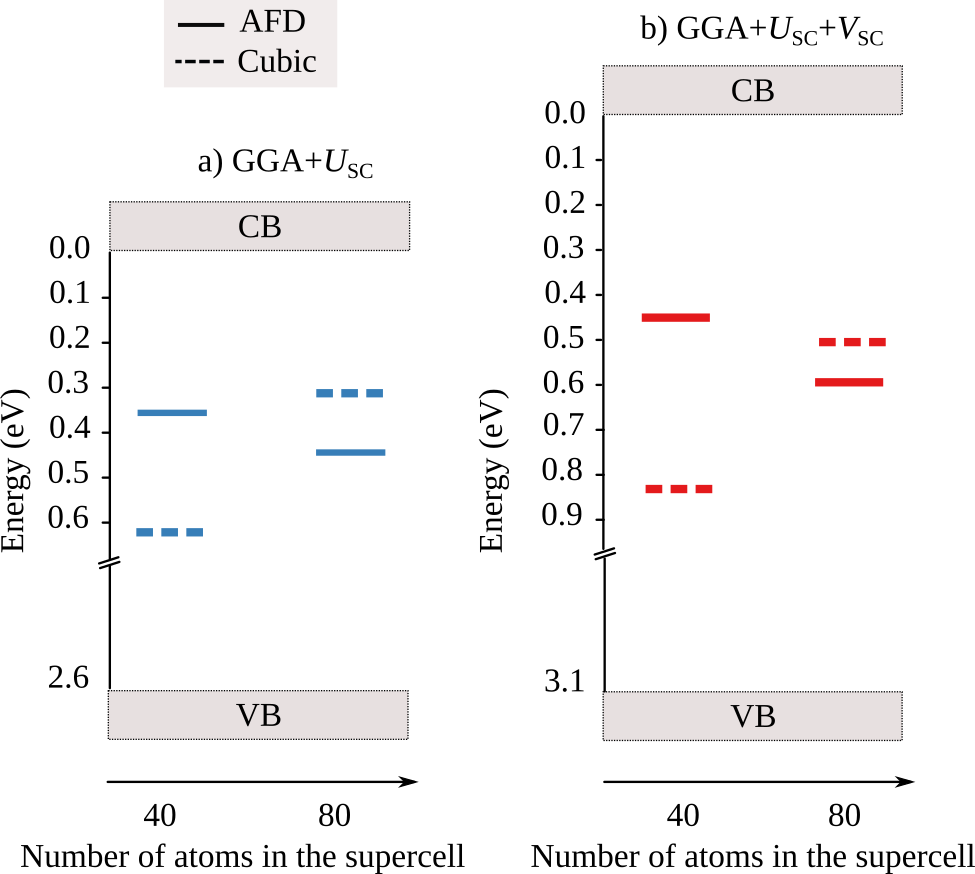}
 \caption{Schematic illustration of the defect-state position of a V$_\textrm{O}^{\bullet}$ for different cell sizes obtained using a) GGA+$U_{\textrm{SC}}$ and b) GGA+$U_{\textrm{SC}}$+$V_{\textrm{SC}}$ in both the cubic (dashed lines) and AFD (solid lines) phases of STO. In both cases, the zero of the energy scale is set to the computed CB minimum.}
\label{fig:defpos_scVO}
\end{figure}

When a neutral oxygen vacancy (V$_\textrm{O}^{\bullet \bullet}$) is formed,  two electrons formerly associated with the O$^{2-}$ anion are left in the lattice. Singly (V$_\textrm{O}^{\bullet }$) and doubly (V$_\textrm{O}^{X}$) positively charged oxygen vacancies correspond, instead, to the removal of a O$^{-}$ and an O$^{2-}$ ion, respectively, thus resulting in one or no extra electron left in the system upon defect formation. Figure~\ref{fig:pdos_charged}a shows the density of states for a V$_\textrm{O}^{\bullet}$ defect, computed at different levels of theory. The underestimation of the band gap at the GGA level, results in the extra electron occupying the bottom of the CB. The larger band gaps  predicted by DFT+$U$ and DFT+$U$+$V$ result, instead, in the appearance of a singly occupied in-gap state, the position of which becomes deeper when going from DFT+$U$ to DFT+$U$+$V$ (\textit{i.e.} with increasing band gap, as schematically shown in Fig.~\ref{fig:defpos_scVO}). When no extra electrons are left in the lattice upon V$_\textrm{O}^{X}$ formation, all methods provide a similar description of the electronic properties, characterized by an empty defect state merged with the CB (Fig.~\ref{fig:pdos_charged}b).
\newpage
\subsubsection{Formation energies for charged defects}
\begin{figure}[h]
 \centering
 \includegraphics[width=0.4\textwidth]{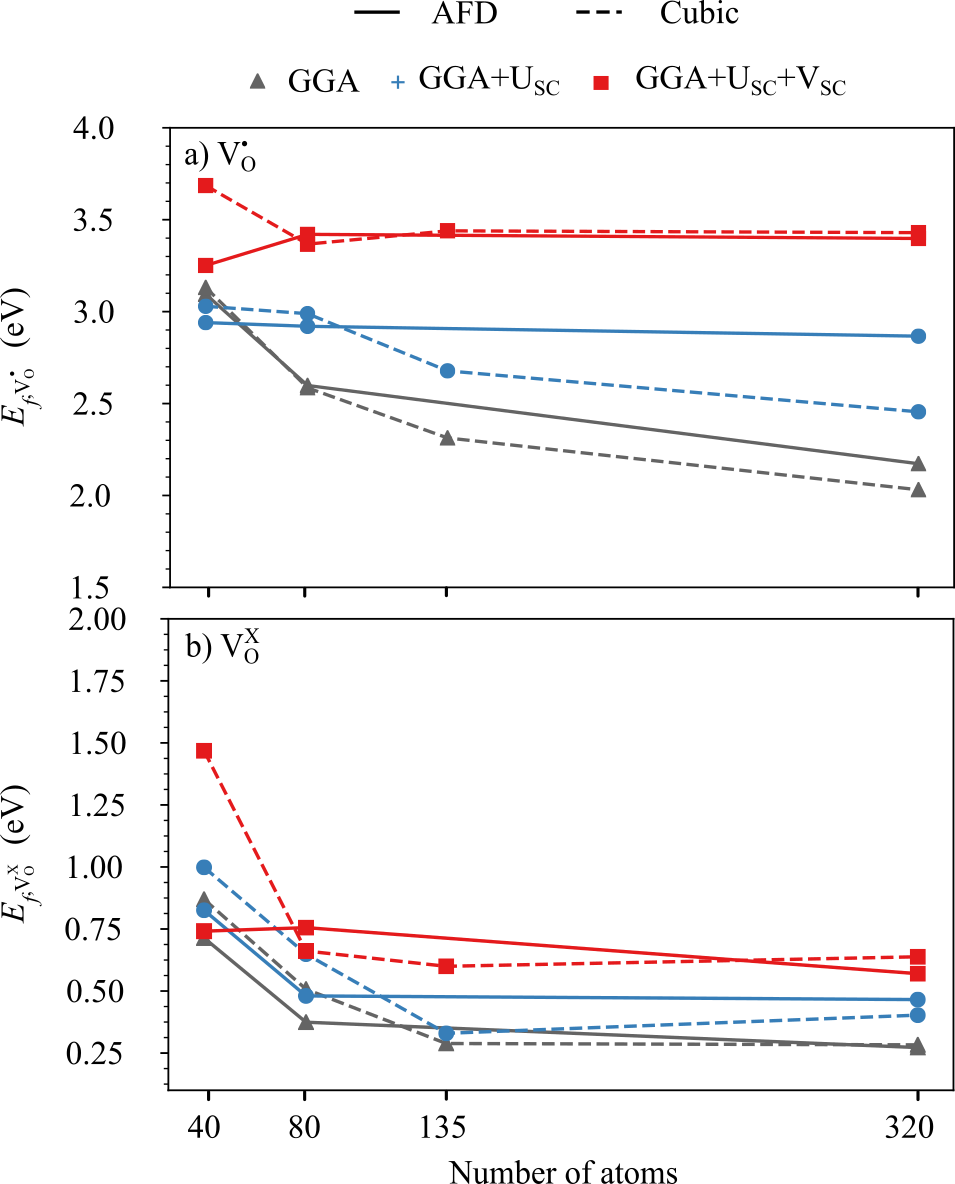}
 \caption{Formation energy for a) a V$_\textrm{O}^{\bullet}$ and b) a V$_\textrm{O}^{X}$ as a function of the cell size in the cubic and AFD phases of STO and computed with different methods. }
\label{fig:ef_charged}
\end{figure}

The change in formation energy of charged defects as a function of the supercell size reflects the different electronic properties as discussed in Sec.~\ref{sec:elecprop_chargedVO}. For the singly charged defect, the formation energy decreases when going from the 40- to the 320-atom cell for the GGA and GGA+$U$ methods due to the strong underestimation of the band gap (see Fig.~\ref{fig:ef_charged}a). However, this effect is smaller than for the neutral oxygen vacancy (see Fig.~\ref{fig:ef_nVO_cellsize} in the main text), in agreement with the fact that upon V$_\textrm{O}^{\bullet}$ formation only one extra electron left in the lattice. Instead, the formation energy stays almost constant when DFT+$U$+$V$ is used, reflecting the improved prediction of the electronic structure provided by this method. Finally, the cell-size dependence of the formation energy is further reduced for the doubly charged defect, as expected due to the electronic structure of  V$_\textrm{O}^{X}$, for which no electron resides in the defect band (see Fig.~\ref{fig:ef_charged}b). 

\newpage
\subsection{Self-Consistent Site-Dependent $U$ and $V$ values}

\subsubsection{Computational details of the site-dependent calculations}\label{sec:methods_scsd}\label{sec:compdetDFPT}

In DFT+$U$ calculations, site-dependent $U$ parameters can be easily computed by perturbing all inequivalent Ti sites in the defective structure identified according to their distance from the defect and their chemical environment (changes in coordination number and/or oxidation state). A similar approach could be applied to the determination of $U$ parameters on the Ti sites also in DFT+$U$+$V$. However, for the determination of $V$ parameters, one needs to consider that O sites are already symmetry inequivalent in the stoichiometric AFD phase, implying that more than one O atom needs to be perturbed. The description becomes much more complex when an oxygen vacancy is created because of the additional symmetry breaking induced by the O removal. If $V$ for the Ti-O pairs in the stoichiometric material can be described by a global value ($V_\mathrm{SC}$), site-dependence should also be taken into account for defects ($V_\mathrm{SC-SD}$) by perturbing an adequate number of Ti and O atoms. In order to simplify and automate these calculations, the atoms to be perturbed were selected based on the unperturbed atomic occupations of Ti and O sites using a difference threshold of 10$^{-6}$. 


\subsubsection{Effect of the cell size on the site-dependent Hubbard parameters}

\begin{figure}[h]
 \centering
 \includegraphics{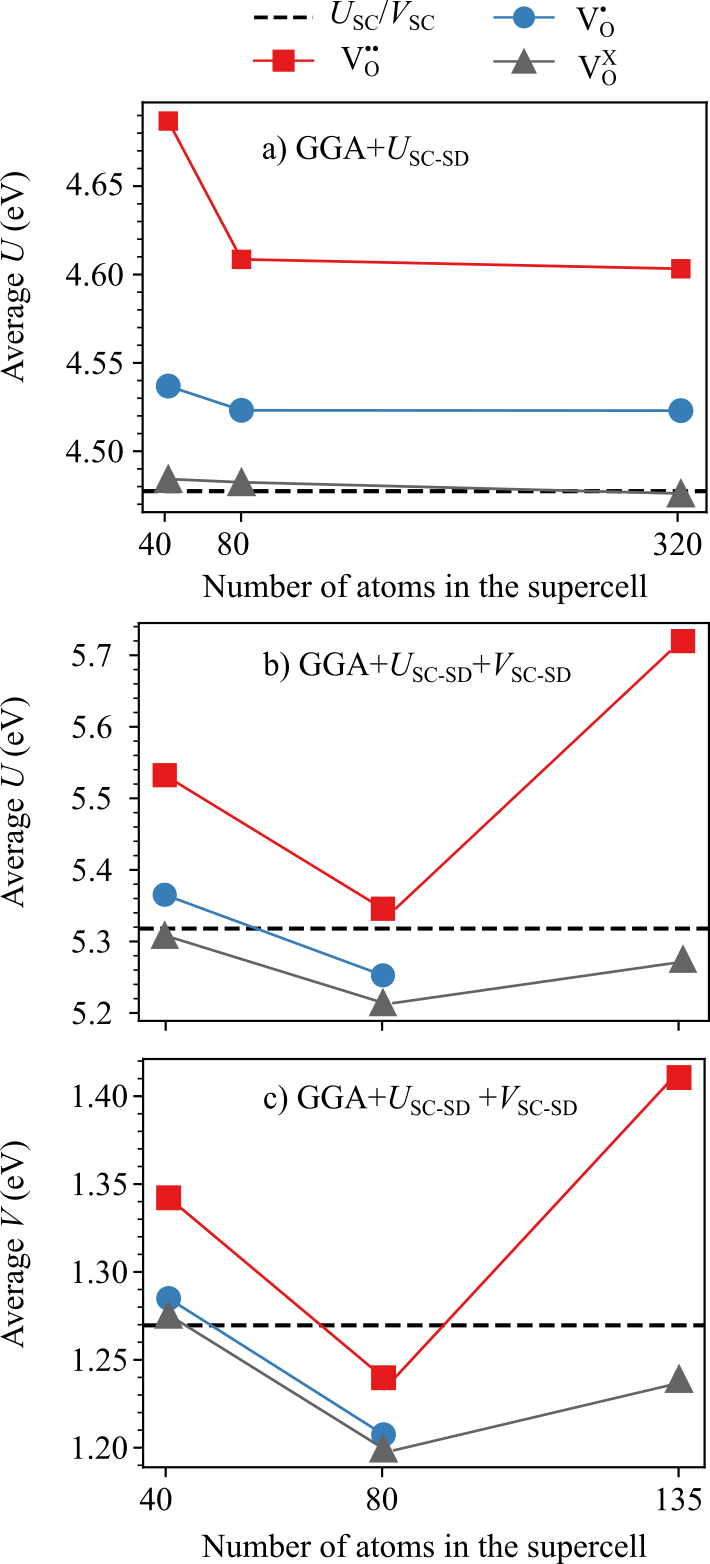}
 \caption{Averaged $U_{\textrm{SC-SD}}$ on all Ti sites  a) for GGA+$U$ and b) for  GGA+$U$+$V$ calculations together with c) the averaged $V_{\textrm{SC-SD}}$ for all Ti(3$d$)-O(2$p$) pairs for AFD supercells of different sizes and containing one V$_{\textrm{SC-SD}}^{\bullet \bullet}$. Dashed horizontal lines indicate the respective SC Hubbard parameters of stoichiometric STO.}
\label{fig:UV_scsd_size}
\end{figure}

Figure~\ref{fig:UV_scsd_size} reports the changes in the $U_{\textrm{SC-SD}}$ and $V_{\textrm{SC-SD}}$ parameters averaged over all Hubbard sites (for $U_{\textrm{SC-SD}}$) or Hubbard Ti-O  pairs (for $V_{\textrm{SC-SD}}$) for  different cell sizes containing one V$_{\textrm{O}}^{\bullet \bullet}$. For DFT+$U_{\textrm{SC-SD}}$+$V_{\textrm{SC-SD}}$ calculations, we did not perform the site-dependent determination of the Hubbard parameters for the largest 320-atom cell as these calculations proved too expensive due to the large number of perturbed Ti and O atoms required for converged results (see Sec.~\ref{sec:compdetDFPT}).

As discussed in the main text for the 40-atom cell, the average value of the Hubbard parameters for a V$_{\textrm{O}}^{\bullet \bullet}$ in Fig.~\ref{fig:UV_scsd_size} is  larger than the one of the stoichiometric cell. This is related to the defect state lying very close to and slightly overlapping with the CB, which results in a non-physical long-range dependence of the site-dependent Hubbard parameters. One would expect that increasing the size of the simulation cell could alleviate this issue due to the reduction of the defect band's dispersion. Unfortunately, this effect is only observed when going from the 40- to the 80- atom cell: the average $U_{\textrm{SC-SD}}$ and $V_{\textrm{SC-SD}}$ are indeed getting closer to the Hubbard parameters of the stoichiometric system for the 80-atom case. However, for larger supercells, the average Hubbard parameters are similar or even larger than the ones obtained for the 80-atom cell, the reduction of the dispersion of the defect state being compensated by its increasing shallowness.  

\clearpage
\bibliographystyleSI{apsrev4-1}
\bibliographySI{references}

\end{document}